\documentclass{emulateapj}
\def\lsim{\raise0.3ex\hbox{$<$}\kern-0.75em{\lower0.65ex\hbox{$\sim$}}}
\def\gsim{\raise0.3ex\hbox{$>$}\kern-0.75em{\lower0.65ex\hbox{$\sim$}}}

\begin{document}

\title{Molecular Outflows From the Protocluster, Serpens South}

\author{Fumitaka Nakamura\altaffilmark{1,2},  
Koji Sugitani\altaffilmark{3},
Yoshito Shimajiri\altaffilmark{4}, 
Takashi Tsukagoshi\altaffilmark{5},  
Aya E. Higuchi\altaffilmark{1}, 
Shogo Nishiyama\altaffilmark{6}, 
Ryohei Kawabe\altaffilmark{4},
Michihiro Takami\altaffilmark{7},
Jennifer L. Karr\altaffilmark{7},
Robert A. Gutermuth\altaffilmark{8},
Grant Wilson\altaffilmark{8}
} 
\altaffiltext{1}{National Astronomical Observatory, Mitaka, Tokyo 181-8588, 
Japan; fumitaka.nakamura@nao.ac.jp}
\altaffiltext{2}{Institute of Space and Astronautical Science, 
Japan Aerospace Exploration Agency, 3-1-1 Yoshinodai, Sagamihara, 
Kanagawa 229-8510, Japan}
\altaffiltext{3}{Graduate School of Natural Sciences, 
Nagoya City University, Mizuho-ku, Nagoya 467-8501, Japan}
\altaffiltext{4}{Nobeyama Radio Observatory, Nobeyama, 
Minamimaki, Minamisaku, Nagano, 384-1305, Japan}
\altaffiltext{5}{Department of Astronomy, 
School of Science, University of Tokyo, Bunkyo,
Tokyo 113-0033, Japan}
\altaffiltext{6}{Department of Astronomy, Kyoto University, Sakyo-ku, Kyoto 606-8502, Japan}
\altaffiltext{7}{Academia Sinica Institute of Astronomy and
Astrophysics, P.O. Box 23-141, Taipei 106, Taiwan}
\altaffiltext{8}{Department of Astronomy, University of Massachusetts, 
Amherst, MA 01003, USA}
\begin{abstract}
We present the results of CO ($J=3-2$) and HCO$^+$ ($J=4-3$)
mapping observations toward a nearby embedded cluster, 
Serpens South, using the ASTE 10 m telescope.  
Our CO ($J=3-2$) map reveals that many outflows
are crowded in the dense cluster-forming clump that can be recognized 
as a HCO$^+$ clump with 
a size of $\sim$ 0.2 pc and mass of $\sim$ 80 M$_\odot$.
The clump contains several subfragments with sizes 
of $\sim$ 0.05 pc.
By comparing the CO ($J=3-2$) map with the 1.1 mm dust continuum 
image taken by AzTEC on ASTE,  
we find that the spatial extents of the outflow lobes are sometimes
anti-correlated with the distribution of the dense gas and 
some of the outflow lobes apparently collide with the dense gas.
The total outflow mass, momentum, and energy are estimated at
0.6 $M_\odot$, 8 $M_\odot$ km s$^{-1}$, and 
64 $M_\odot$ km$^2$ s$^{-2}$, respectively.
The energy injection rate due to the outflows is 
comparable to the turbulence dissipation rate 
in the clump, implying that the protostellar outflows can
maintain the supersonic turbulence in this region.
The total outflow energy seems only about 10 percent the 
clump gravitational energy. 
We conclude that the current outflow activity is not enough
to destroy the whole cluster-forming clump, and therefore 
star formation is likely to continue for several or many 
local dynamical times.
\end{abstract}

\section{Introduction}
\label{intro}

Most stars form in clusters \citep{lada03,allen06}. 
Therefore, understanding the formation process of star clusters is 
a key step towards a full understanding of how stars form.
Recent observations have revealed that star clusters form 
in turbulent, magnetized, parsec-scale dense clumps of molecular clouds
 \citep{ridge03}.
These clumps contain masses of $10^2-10^3 M_\odot$,
fragmenting into an assembly of cores 
that collapse to produce stars
\citep[e.g.][]{davis99,sandell01,peretto06,andre07,kirk07,
walsh07,saito08,maruta10}. 
In cluster-forming clumps, stellar feedback such as protostellar outflows, 
stellar winds, and radiation rapidly start to shape 
the surroundings. 
Because of the short separations between forming stars 
and cores, these feedback mechanisms are expected to control 
subsequent star formation 
\citep[e.g.,][]{norman80,krumholz06,mckee07,matzner07,
nakamura07,peters10,wang10}. 
However, the roles of the stellar feedback on cluster formation 
remain poorly understood observationally.

According to recent theoretical studies, two main 
scenarios of cluster formation are now under debate. 
For these two scenarios, 
the roles of the stellar feedback on cluster formation are 
different.
One is the rapid, dynamical formation model, in which 
the large-scale turbulent flow mainly controls star formation in
cluster-forming clumps, and star formation is envisioned to 
complete in a short time \citep[e.g.,][]{elmegreen07,hartmann07}. 
In this case, cluster formation is expected to be 
terminated by stellar feedback due to initial star burst 
within a couple of turbulence-crossing times, so that 
star formation efficiencies (SFEs) can stay low at a observed level in 
nearby cluster-forming regions (SFE $\sim $ 0.1).
To promote rapid gravitational collapse and thus rapid star formation,
the magnetic fields should  be weak dynamically.
Another scenario is the quasi-virial equilibrium model, in which 
cluster formation is considered to continue at least for several
dynamical, or free-fall, times \citep[e.g.,][]{tan06,li06}.
Because supersonic turbulence dissipates rapidly
in a turbulence-crossing time, additional turbulent motions
should be injected to maintain the supersonic turbulence.
Protostellar outflow feedback is proposed to play a significant 
role in the turbulence regeneration  
\citep{matzner07,nakamura07,carroll10}.
The moderately-strong magnetic fields are also important
to impede the rapid global gravitational collapse.
In this case, the global inflow and outflow are expected to coexist,
interacting with themselves. As a result, the cluster-forming clumps 
as a whole can keep quasi-equilibrium states for a relatively long time.

In an effort to discriminate between the two scenarios
and constrain the theory of cluster formation,
we have made a start on a systematic observational study
of a nearby infrared dark cloud (IRDC), Serpens South,
with a focus on the stellar feedback.
Recent observations strongly suggest that 
IRDCs are in the very early phase of star cluster formation 
\citep{rathborne06, perreto09, butler09} and presumably 
still retain the primordial structure of the cluster's natal cloud.
In particular, in the early phase of evolution,
the protostellar outflows are likely to play a more dominant role
in cloud dynamics, compared to the stellar winds and radiation, 
because they are more powerful in the early protostellar evolution
\citep{bontemps96}.
Therefore, IRDCs are expected to be suitable to explore
the roles of protostellar outflows on cluster formation.

Serpens South is a nearby embedded cluster,
recently discovered by \citet{gutermuth08} using 
the Spitzer Space telescope. 
The accurate distance to Serpens South remains unknown. Most of 
the previous studies adopt the same distance as 
the Serpens Cloud Core, located about 3$^\circ$ north, 
because the YSOs associated with Serpens South have
the same LSR velocities  as the Serpens Cloud Core, 
whose distance is usually adopted 
as 260 $\pm$ 37 pc \citep{straizys96}.  
However, larger distances of 415$\pm 25$ pc, 
380 pc, and 360$^{+22}_{-13}$ pc have been recently claimed 
for the Serpens Cloud Core based on a VLBA
parallax of EC95, a young AeBe star embedded in Serpens 
Cloud Core \citep{dzib10}, SEDs of YSOs 
base on the near infrared observations \citep{gorlova10}, 
and X-ray luminosity functions
\citep{winston10}, respectively.  
According to \citet{gutermuth08}, the Serpens South cloud
appears to be seen in absorption against PAH emission from the W40
HII region.  Therefore, the distance to W40 
is likely to be an upper limit to the distance to Serpens South, 
although even the former has not yet been determined 
to any satisfactory precision
\citep[300 pc $-$ 900 pc; ][]{rodney08}.
In this paper, we assume 260 pc as the distance to Serpens South.

From the Spitzer IRAC observations, 
\citet{gutermuth08} revealed that the cluster contains 
about 60 Class I protostars and 40 Class II sources.
Recently, \citet{bontemps10}
have discovered 7 Class 0 candidates in this region through
the Herschel Gould Belt Survey \citep[see][]{andre10}. 
In the central part of the cluster, 
the fraction of Class I protostars relative 
to all the YSOs (Class I/II) reaches about 80 \% at the high 
surface density of 430 pc$^{-2}$. 
These observations indicate that Serpens South is 
undoubtfully in the very early stage of cluster formation, 
i.e., Serpens South is a ``protocluster'' 
that are forming a number of stars.
\citet{gutermuth08} also estimated the median projected distance 
between nearest neighbor YSOs to be about 3700 AU, significantly shorter 
than the typical length of protostellar outflow lobes,
implying that the outflows can potentially influence
star formation in this region.

Recent dust continuum observations have revealed that 
the cluster is located at the constricted region 
in a long filamentary cloud that is recognized as a dark cloud 
in mid-infrared ($\sim 10 \mu$m) wavelength
\citep{gutermuth08,gutermuth11,andre10}. 
Several sub-filaments also appear to converge toward 
the main filament \citep[e.g.][]{myers09,bontemps10}.
Very recently, \citet{sugitani11} carried out near-infrared
(JHKs) polarization observations toward Serpens South using 
the imaging polarimeter SIRPOL which is a polarimetry mode
of the near-infrared camera SIRIUS mounted on the IRSF 1.4 m 
telescope at the South Africa Astronomical Observatory.
They found that the global magnetic field is spatially well-ordered
and almost perpendicular to the main filament. Several sub-filaments
appear to be parallel to the global magnetic field.
These observational facts imply that moderately-strong magnetic field
played a role in the course of the formation of the Serpens 
South filament.  
The moderately-strong magnetic field 
appears consistent with the quasi-virial equilibrium model
where the magnetic fields significantly impede the global
gravitational collapse \citep[see, e.g.,][]{nakamura11b}.
In this paper, we investigate the outflow activity in 
the Serpens South cloud on the basis of the CO ($J=3-2$) and
HCO$^+$ ($J=4-3$) observations using the ASTE 10 m telescope.

The rest of the paper is organized as follows. First, we
describe the details of our CO $(J=3-2)$ and HCO$^+$ $(J=4-3)$
observations toward Serpens South in Section \ref{sec:observations}.
We present in Section \ref{sec:results} our CO $(J=3-2)$ and 
HCO$^+$ $(J=4-3)$ maps toward Serpens South, which reveal a number of 
powerful collimated outflows and clumpy structure in the central part 
of the protocluster, respectively.  We also compare our CO $(J=3-2)$ map with 
the three-color Spitzer IRAC image (3.6 $\mu$m in blue, 
4.5 $\mu$m in green, 8.0 $\mu$m in red)
to help identify the molecular outflow lobes.
In the three-color image, the knots created by shocks due to 
the outflows are found as extended green objects 
with excess in 4.5 $\mu$m emission
\citep[e.g.,][]{takami10}.
We derive the global properties of the outflows identified 
from our CO $(J=3-2)$ map.
Then, in Section \ref{sec:discussion}, we briefly discuss
how the observed outflows influence the clump dynamics.
Finally, we summarize our main conclusion in Section \ref{sec:summary}.

\section{Observations}
\label{sec:observations}

\begin{figure}
\plotone{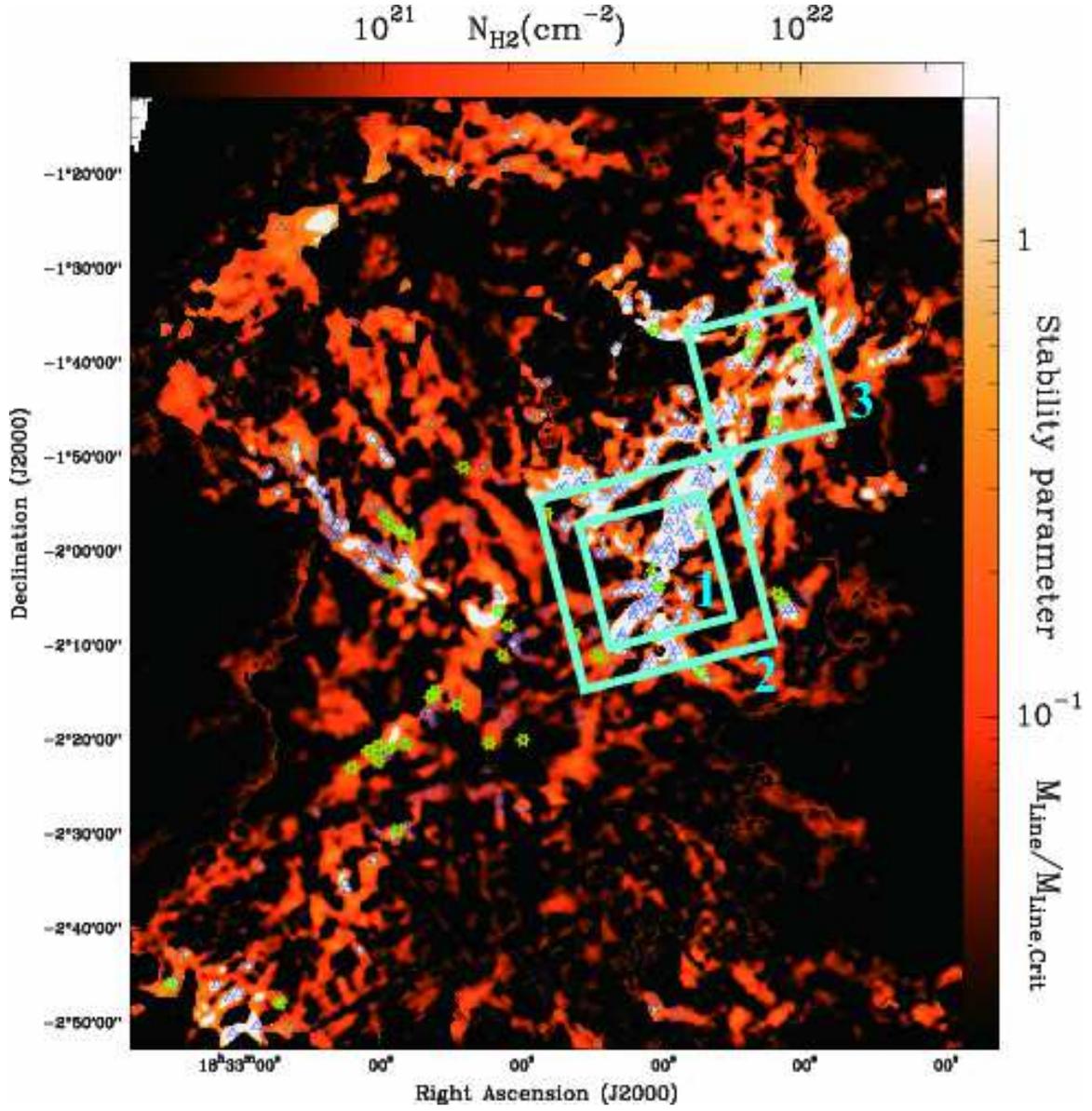}
\caption{Our observation area overlaid on 
the column density map derived from the Herschel
observations.  The image is taken from 
the left panel of Fig. 1 of \citet{andre10}, in which 
the contrast of the filaments was enhanced by a curvelet
transform (see the Appendix of Andr{\'e} et al. 2010).
The blue triangles and green stars indicate the prestellar cores and
 candidate Class 0 protostars identified with Herschel, respectively.
Our observation area is indicated in cyan boxes
that are labeled in numbers.
}  
\label{fig:obsarea}
\end{figure}

The CO ($J=3-2$; 345.796GHz) and HCO$^+$ 
($J=4-3$; 356.734GHz) mapping observations were carried out 
with the ASTE 10 m telescope \citep{ezawa04} toward Serpens South 
during the period of August 12 $-$ 14 2010, using 
the on-the-fly (OTF) mapping technique (Sawada et al. 2008). 
Figure \ref{fig:obsarea} presents the observation boxes
that are overlaid on the column density map derived from 
the Herschel observations \citep{andre10}, indicating that 
our observation area covers most of the dense filamentary structure
associated with Serpens South.
Our main observation box (Box 1 in Fig. \ref{fig:obsarea}) 
covers a $13'\times 13'$ region centered at the dense part of the
protocluster, (RA, Dec) = (18h30m3.4s, $-2^\circ$2$'$4.6$''$). 
We make the image, by combining scans along the two axes 
that run at right angles to one another  
using the PLAIT algorithm developed by 
Emerson \& Graeve (1988), in order to minimize 
the ``scanning effect''.
The observation region was extended to a larger
 $20'\times 20'$ box (Box 2 in Fig. \ref{fig:obsarea}). 
We also observed  a $13'\times 13'$ box 
(Box 3 in Fig. \ref{fig:obsarea}) centered on 
(RA, Dec) = (18h29m16s, $-1^\circ$41$'$35$''$). 
The extended observation areas (Boxes 2 and 3) have higher noises and 
suffer from the scanning effects due to the OTF observations because 
only one scan was done.
The beam size in HPBW of the ASTE telescope is 22$''$, which corresponds 
to 0.028 pc at a distance of 260 pc.
The main-beam efficiency, $\eta$, was 
0.57 at 345 GHz.
We used a 345 GHz SIS heterodyne receiver, which had the 
typical system noise temperature of about 200 K in DSB mode 
at the observed elevation. 
The temperature scale was determined by the chopper-wheel method, 
which provides us with the antenna temperature, $T_A^*$, corrected 
for the atmospheric attenuation. As a backend, we used four sets 
of 1024 channel autocorrelators, providing us with a frequency 
resolution of 125 KHz that corresponds to 0.11 km s$^{-1}$
at 345 GHz. 
After subtracting linear baselines, the data were convolved with a 
Gaussian-tapered Bessel function (Magnum et al. 2007) 
and were resampled onto a $5''$ grid. 
The resultant effective FWHM resolution is $24''$.
The typical rms noise levels are 0.11 K and 0.51 K in $T_{\rm A}^*$ 
with a velocity resolution of 0.5 km s$^{-1}$ 
for Box 1 and Box 2 and 3, respectively. 

As for the Spitzer IRAC data, archival data in three IRAC bands 
(3.6, 4.5, and 8.0 $\mu$m) were obtained for Serpens South.
All the data had been reduced with the basic calibration data 
(BCD) pipelines developed by Infrared Processing and 
Analysis Center (IPAC), and mosaicked with the MOsaicking 
and Point Source Extraction (MOPEX) software developed
by Spitzer Science Center.
The mean FWHMs of the point 
response functions (PRFs) are 1$''$.66, 1$''$.72, 1$''$.98, 
for the three bands, respectively 
(See Spitzer Space Telescope Observer's Manual Version 8.0).

\section{Results}
\label{sec:results}

\subsection{Global Cloud Properties}
\label{subsec:global}

\begin{figure}
\plotone{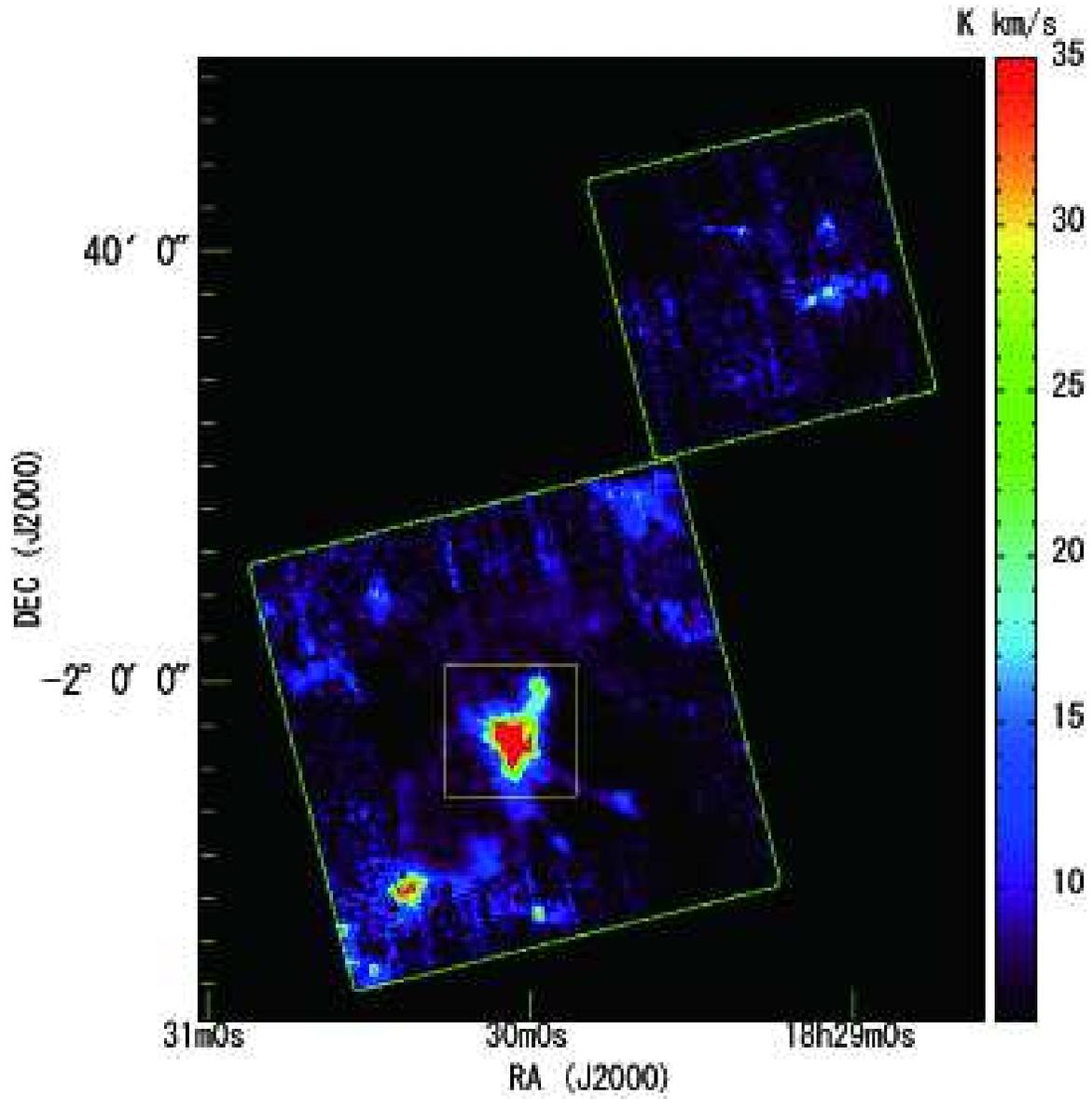}
\caption{CO ($J=3-2$) total integrated intensity map toward Serpens South 
 in the velocity range from $v_{\rm LSR} = -2$ km s$^{-1}$ to +15 km s$^{-1}$.
The blow-up of the yellow box is presented in the upper panel of 
Fig. \ref{fig:profile1}.
}  
\label{fig:copeak}
\end{figure}

Since both the CO ($J=3-2$) and HCO$^+$ ($J=4-3$) transitions have 
higher critical densities  
(approximately, $10^4$ and 10$^7$ cm$^{-3}$, respectively) 
and higher transition energies  
($\sim 33$ and 43 K, respectively) than those of the lower transition 
lines such as CO ($J=1-0$), it allows us to examine higher density 
and/or higher temperature gas. 
In Figure \ref{fig:copeak},
we present the CO ($J=3-2$) 
velocity integrated intensity map 
in the ranges 
from $v_{\rm LSR} = -2$ km s$^{-1}$ to +15 km s$^{-1}$, 
which shows that the CO ($J=3-2$) emission tends to be much stronger 
in the southern box with about $20' \times 20'$ area (Boxes 1 and 2), 
while a relatively weak emission can be seen in the northern box 
with about $13' \times 13'$ area (Box 3).
In fact, the Serpens South protocluster resides in Box 1, 
and the strong CO ($J=3-2$) emission originates mainly from 
the powerful molecular outflows of the cluster member YSOs, 
as described in the next subsection.
We note that the weak CO ($J=3-2$) emission was detected in the entire 
observed area, indicating that the molecular gas is widely distributed 
in much larger area.

\begin{figure}
\epsscale{0.4}
\plotone{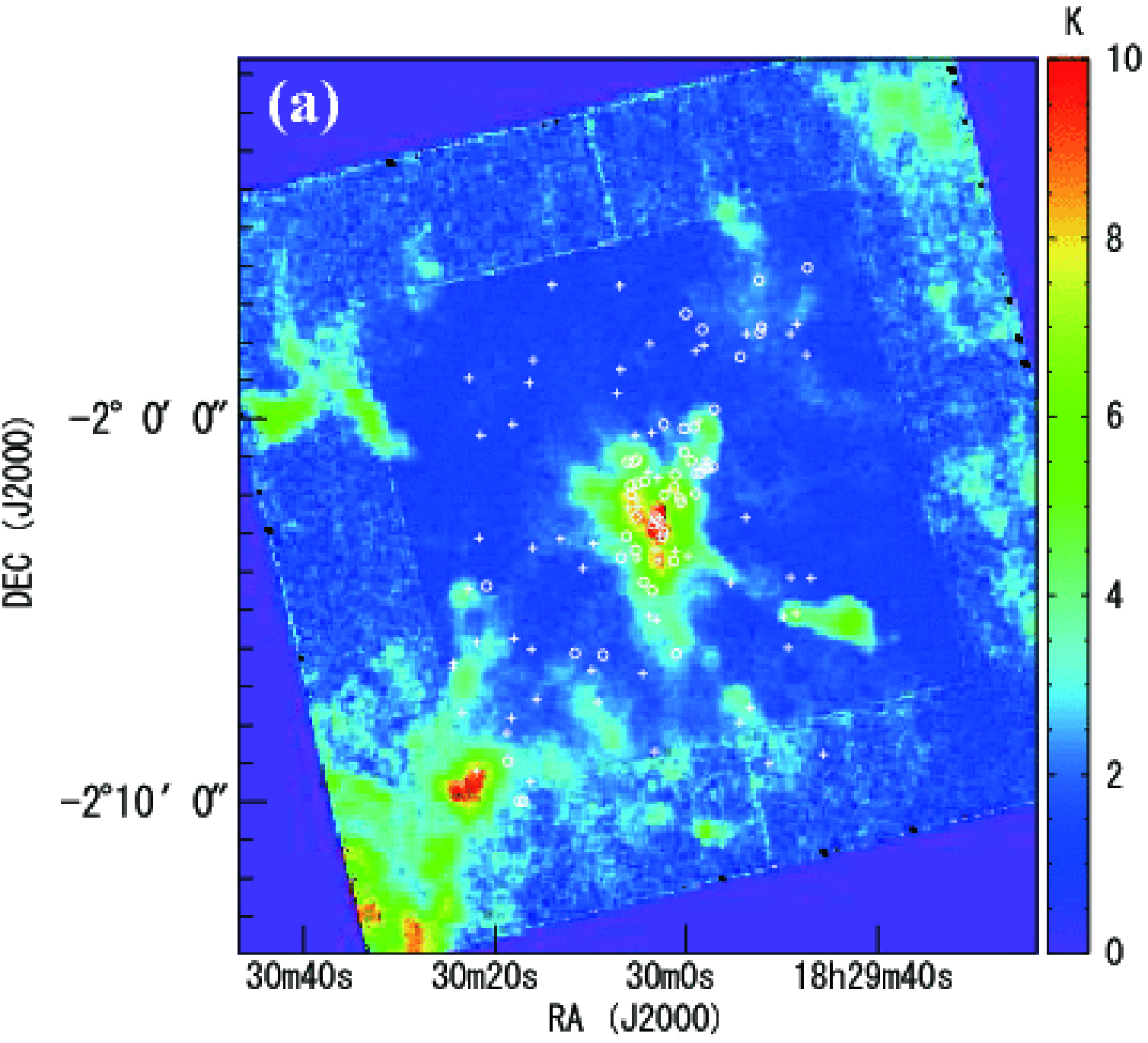}
\plotone{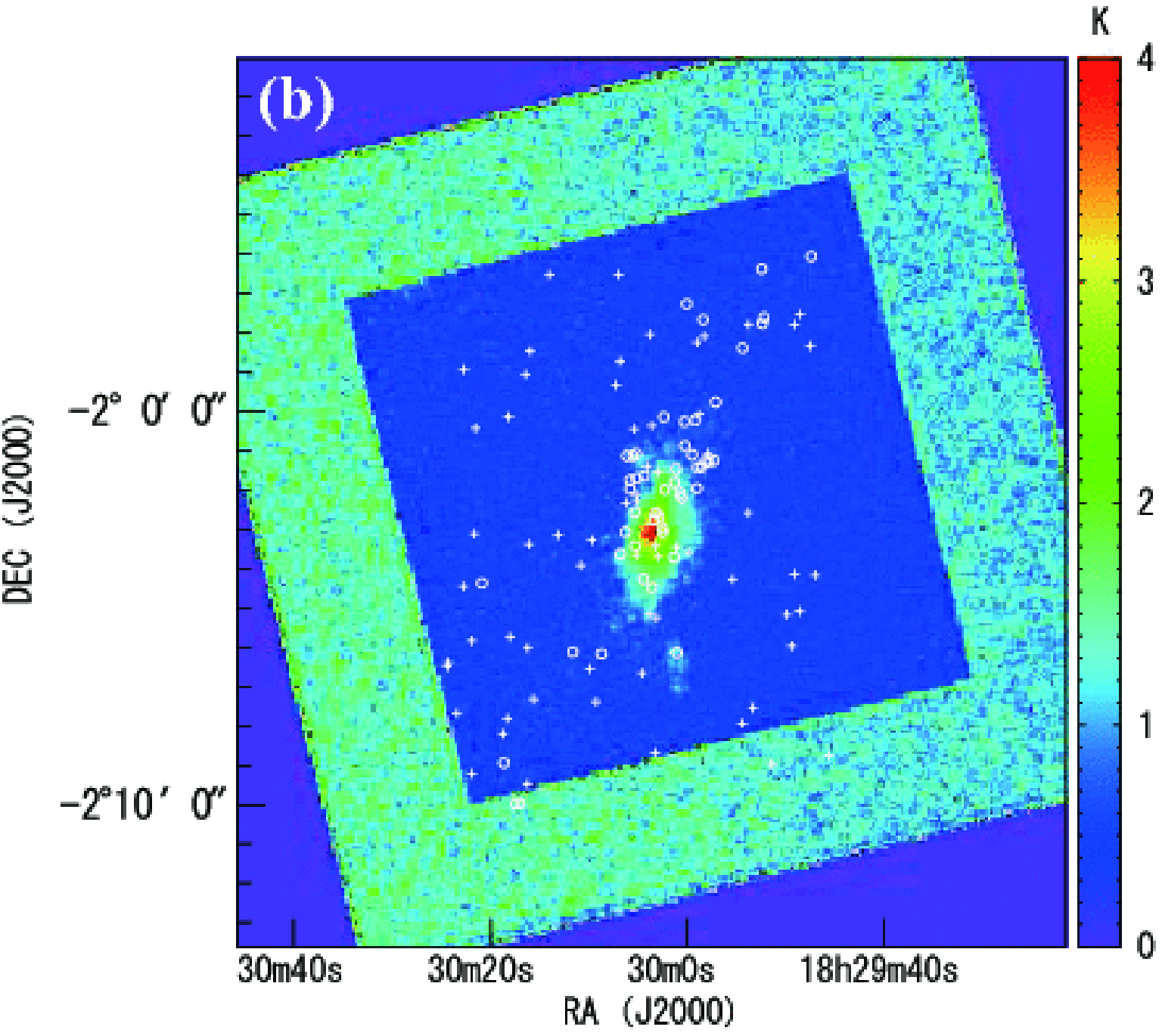}
\plotone{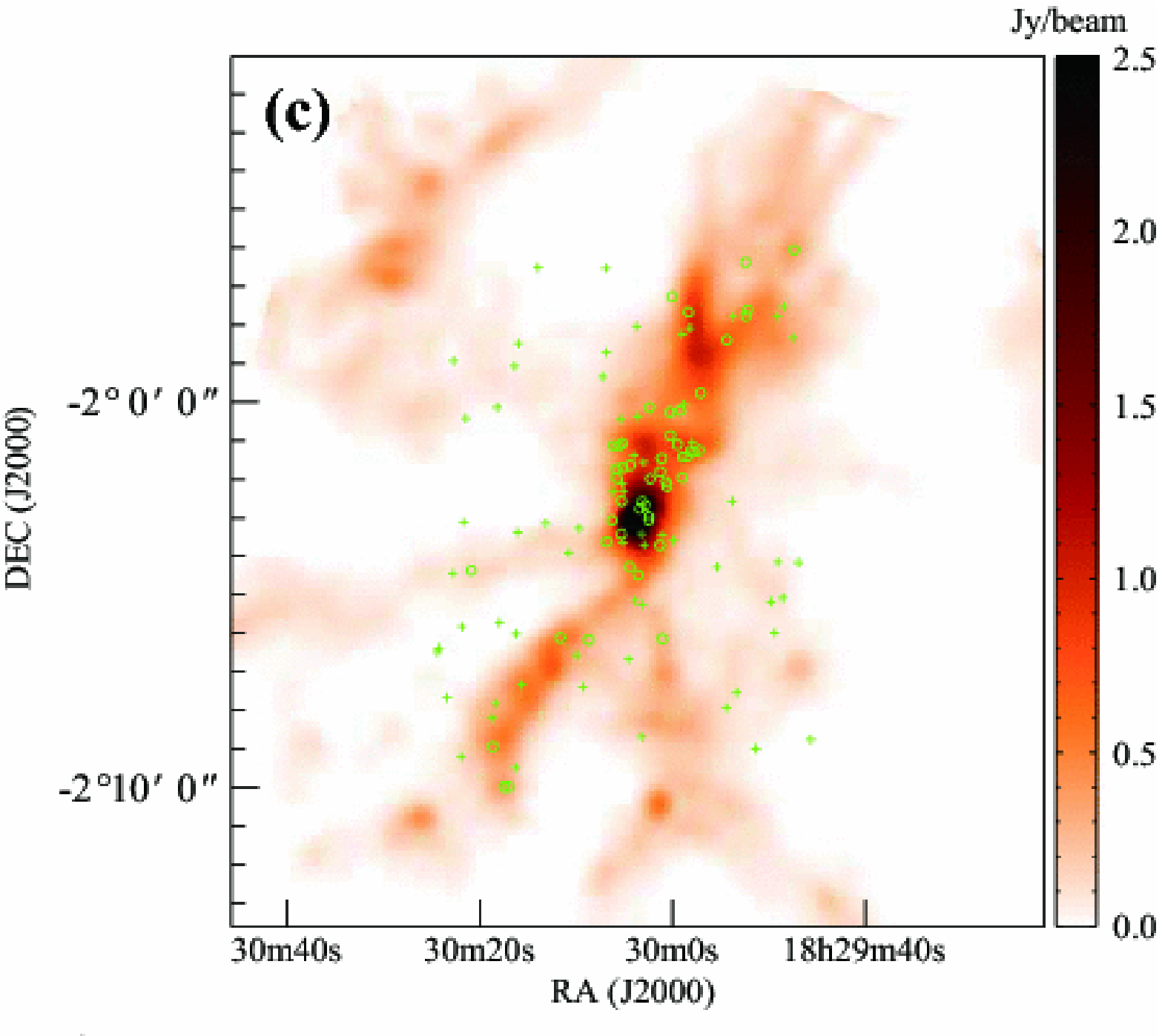}
\caption{(a) CO ($J=3-2$) peak intensity map toward the Serpens South
protocluster.
(b) HCO$^+$ ($J=4-3$) peak intensity map toward the same area
presented in panel (a). 
(c) 1.1 mm continuum emission map toward the same area
presented in panel (a). The detail of the image will be described in
\citet{gutermuth11}.
For all the panels, the circles and crosses indicate the Class I and
 Class II sources, respectively, identified by \citet{gutermuth08}.
}  
\label{fig:peaksmall}
\end{figure}

We also present in Figures \ref{fig:peaksmall}a, 
\ref{fig:peaksmall}b, and \ref{fig:peaksmall}c, the 
CO ($J=3-2$) peak intensity, HCO$^+$ ($J=4-3$) peak intensity,
and 1.1 mm continuum maps, respectively, toward Box 2. 
The 1.1 mm continuum map was taken by the AzTEC camera 
\citep{wilson08}
on the ASTE 10 m telescope \citep{gutermuth11}.
The positions of Class I and II sources identified
by \citet{gutermuth08} from the Spitzer IRAC observations
are plotted in Figures \ref{fig:peaksmall}a through \ref{fig:peaksmall}c. 
The 1.1 mm map indicates that the dense gas is distributed along a 
long filament running from north to south. This filament
(hereafter, the main filament)
has several sub-filaments that appear to converge toward 
the main filament. Such structures appear to be common
in many star-forming region, as pointed out by \citet{myers09}.

The CO ($J=3-2$) peak intensity map in Fig. \ref{fig:copeak}a 
is not dominated by emission at the systemic velocity of the cloud, 
$\sim$ 7.5 km s$^{-1}$.
Such a feature is prominent in the velocity channel maps presented in 
Figs. \ref{fig:channelmap} and \ref{fig:channelmap2},
 where the emission is weak in the range from $\sim 7$ to $\sim$ 10 km s$^{-1}$.
For comparison, we also present in Figures \ref{fig:profile1}  and 
\ref{fig:profile3} 
the CO ($J=3-2$) and HCO$^+$ ($J=4-3$) spectra, respectively. 
The CO ($J=3-2$) spectra often show double-peak line profiles, whereas 
the HCO$^+$ ($J=4-3$) spectra often show single-peak line profiles.
By comparing the CO ($J=3-2$) and HCO$^+$ ($J=4-3$) spectra,
it is clear that the CO ($J=3-2$) line profiles appear to be caused by 
self-absorption in the range from $\sim 7$ to $\sim$ 10 km s$^{-1}$,
rather than multiple velocity components.

Strong CO ($J=3-2$) and HCO$^+$ ($J=4-3$) emission come from 
almost the same part where the 1.1 mm emission is strongest
and the Serpens South protocluster is located.
In fact, the CO ($J=3-2$) and HCO$^+$ ($J=4-3$) emission
peaks coincide reasonably well with the 1.1 mm peak.
The HCO$^+$ ($J=4-3$) emission is distributed in an 
elongated clump with the axis ratio of about $1.5 - 2$.
The HCO$^+$ ($J=4-3$) integrated intensity map 
presented in Fig. \ref{fig:profile3} suggests that the clump 
contains several subfragments. 
The clumpy structure of the HCO$^+$ clump is more 
evident in the HCO$^+$ ($J=4-3$) velocity channel maps  
presented in Fig. \ref{fig:hcochannelmap}.
The typical size of the subfragments is about 0.05 pc.
The positions of two subfragments (sources 1 and 2 indicated in 
Fig. \ref{fig:spitzer2}b) are indicated by 
the diamonds in Fig. \ref{fig:hcochannelmap}.  
As shown in the next subsection, these two subfragments 
contain the possible driving sources of the outflows.
The HCO$^+$ ($J=4-3$) emission takes its maximum at the position of 
source 2 presented in Fig. \ref{fig:spitzer2} 
that coincides with the position of the 1.1 mm peak.
The positions of these two subfragments coincide 
reasonably well with those of the 3 mm compact sources
recently identified by the NMA observations (Fukuda et al.
in prep.).

The CO line profiles have prominent blueshifted and/or 
redshifted wings near the region where the CO emission is strongest, 
as can be seen in Fig. \ref{fig:profile1}.
The line wings sometimes extend over the velocity range 
of $-10 \sim $ 30 km s$^{-1}$.
These prominent wings indicate the presence of powerful molecular 
outflows in this region.
The CO ($J=3-2$) peak intensity map indicates that there are several local 
peaks and ridges near the central dense part of the protocluster.
As shown in Section \ref{subsec:outflow}, these peaks 
trace the outflow lobes reasonably well.  
Similar high-velocity wings can also been seen in the HCO$^+$ ($J=4-3$) 
line profiles (see Fig. \ref{fig:profile3}) that are   
are sometimes extended in the
range from about 3 km s$^{-1}$ to 15 km s$^{-1}$ around the
subfragments located near the 1.1 mm peak.

In the southern part of Box 2, the redshifted components of the 
identified outflows tend to be extremely weak (see also the 
line profiles presented in Fig. \ref{fig:profile1}b).  
It is unclear what causes the extremely-asymmetric line profiles
in the southern area. One possibility is the existence of a 
foreground cloud in the velocity of around 10 km s$^{-1}$ 
in the southern area. The foreground cloud, if cold, 
preferentially absorbs the redshifted outflow components.  
Another possibility is the existence of the global infall 
toward the southern part of the cloud.  If the cloud envelope 
is infalling, the foreground infall gas having velocities 
similar to the redshifted outflow components can 
absorb the redshifted components.
In the next subsection, we identify the molecular outflow lobes from 
CO ($J=3-2$) data and discuss several features of the outflow 
lobes in detail.

\subsection{Molecular Outflows}
\label{subsec:outflow}

\begin{figure}
\epsscale{1.0}
\plotone{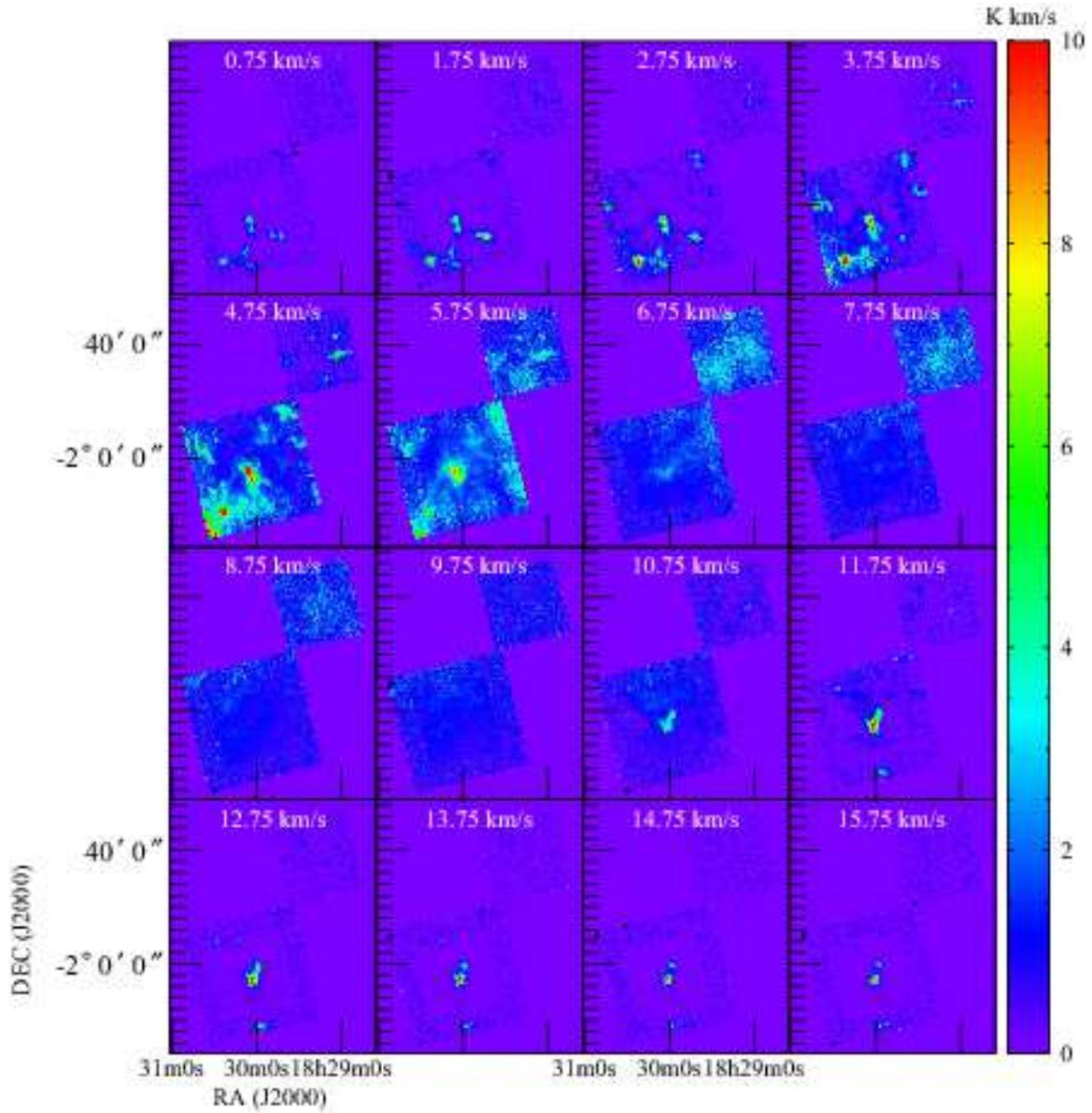}
\caption{
Velocity channel maps of the CO ($J=3-2$) emission
with the velocity intervals of 1 km s$^{-1}$.
The LSR velocity is indicated at the top of each panel.
}  
\label{fig:channelmap}
\end{figure}

\begin{figure}
\epsscale{1.0}
\plotone{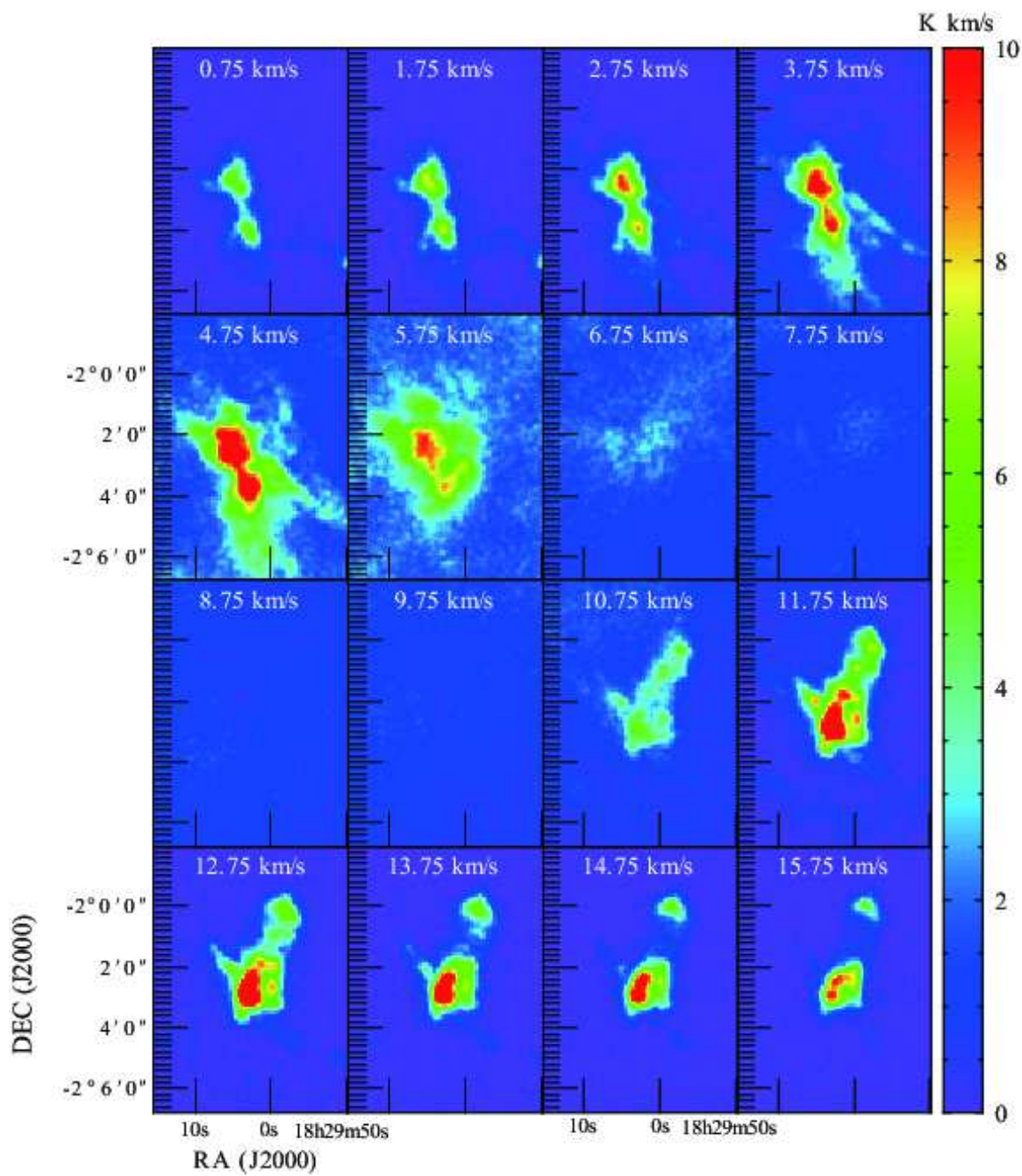}
\caption{Same as Fig. \ref{fig:channelmap} but toward 
the central region of Serpens South.
}  
\label{fig:channelmap2}
\end{figure}

\begin{figure}
\epsscale{0.6}
\plotone{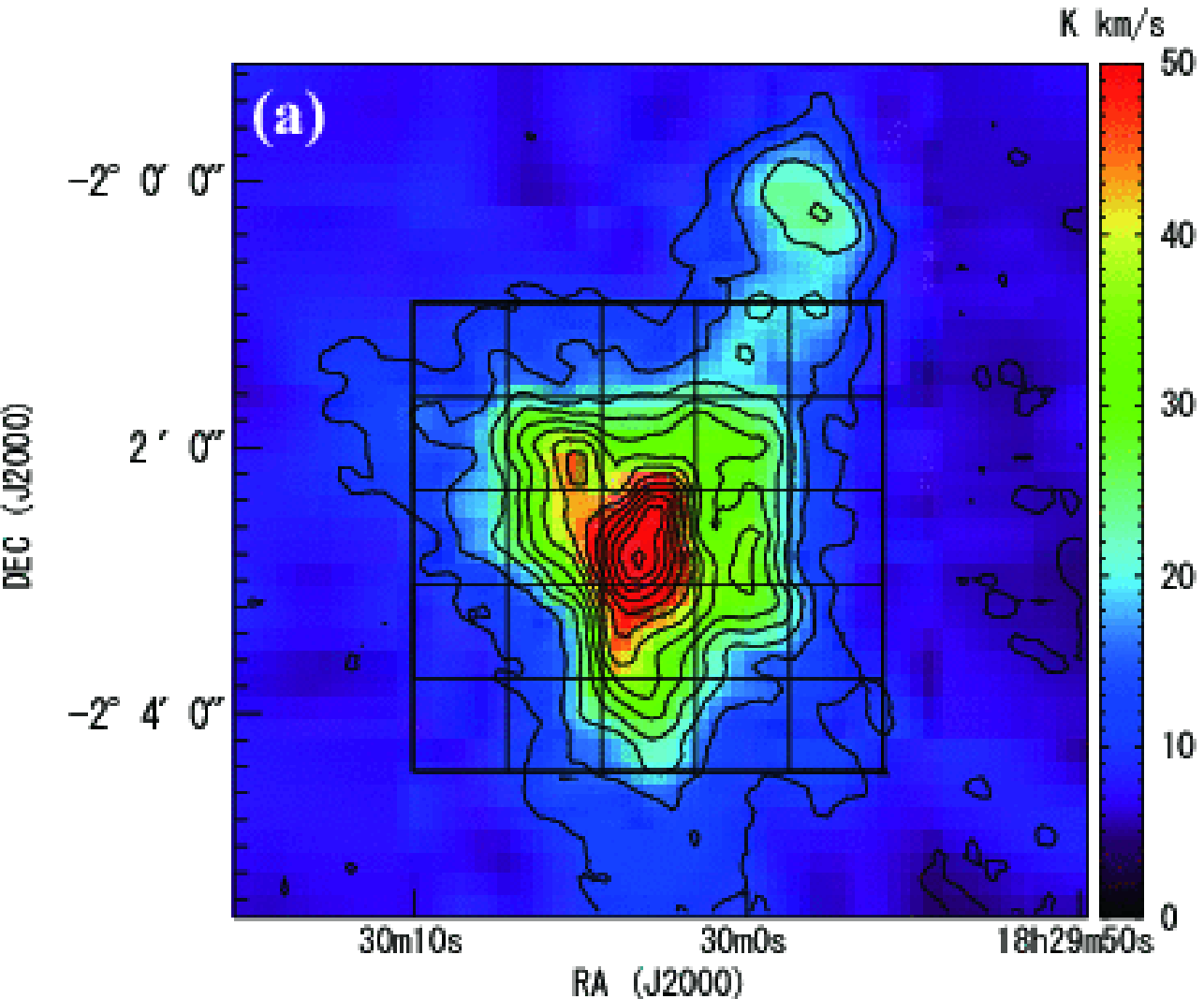}
\plotone{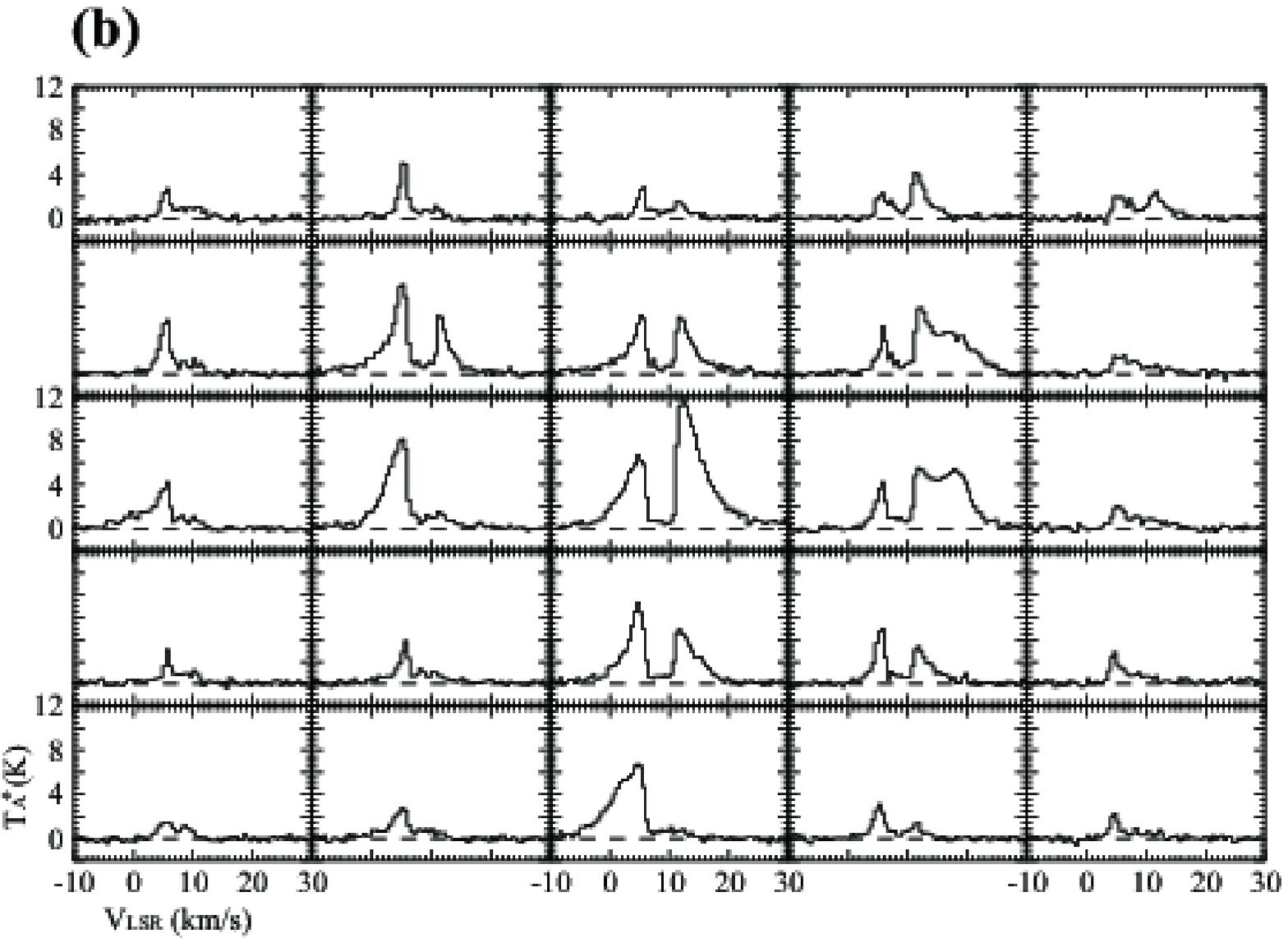}
\caption{{\it upper panel}: CO ($J=3-2$) total integrated intensity map 
 in the velocity
range from $v_{\rm LSR} = -2$ km s$^{-1}$ to +15 km s$^{-1}$
toward the yellow box indicated in Fig. \ref{fig:copeak}.
The CO ($J=3-2$) integrated intensity is shown 
in color in units of K km s$^{-1}$.
The contours start at 5 K km s$^{-1}$ at an interval of 5 K km s$^{-1}$.
The boxes indicate the areas where CO ($J=3-2$) spectra take average
(see the lower panel of Figure \ref{fig:profile1}).
{\it lower panel}: CO ($J=3-2$) spectra. Each profile is averaged inside each 
box indicated in Figure \ref{fig:profile1}a.
}  
\label{fig:profile1}
\end{figure}

\begin{figure}
\epsscale{0.6}
\plotone{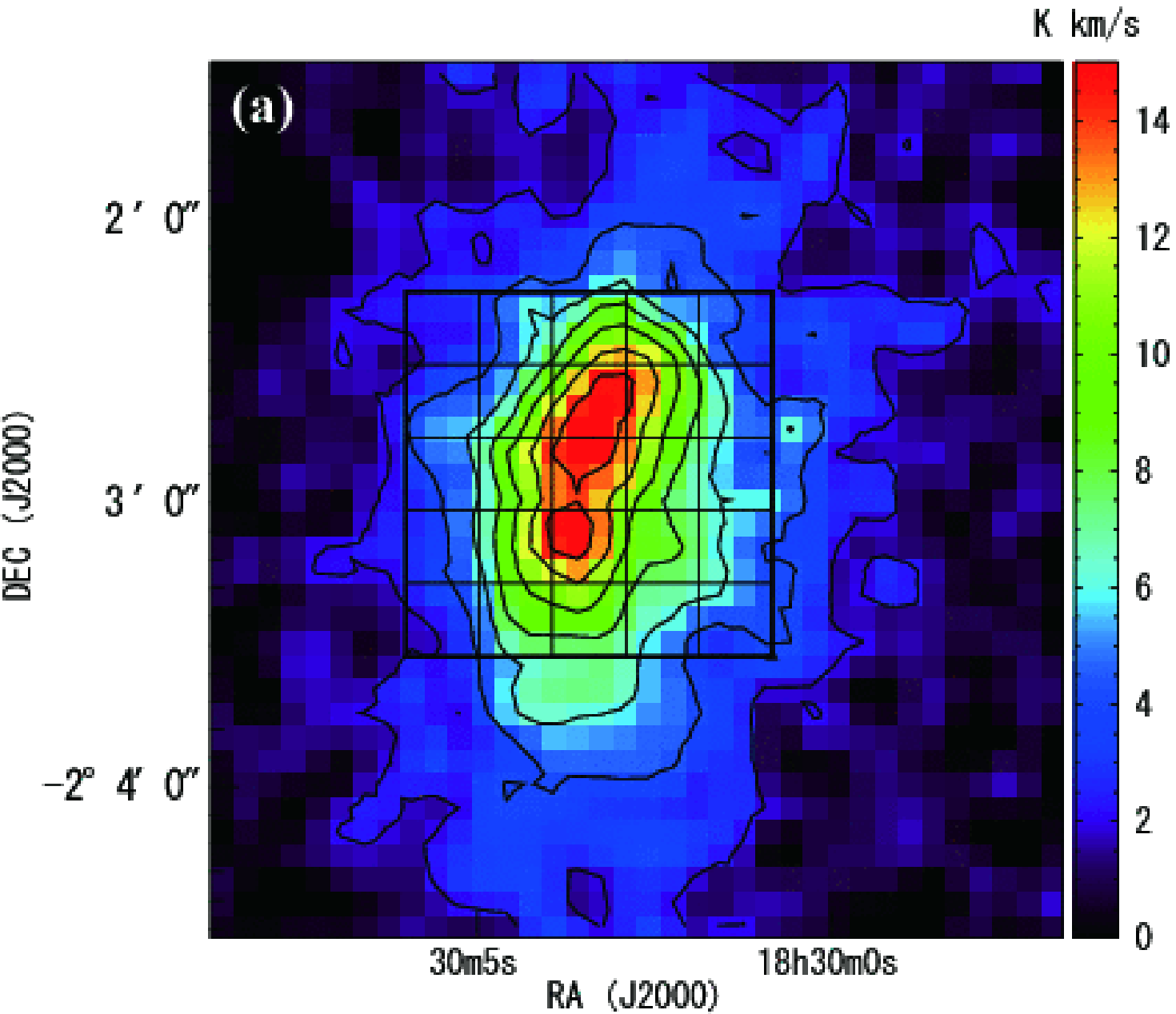}
\plotone{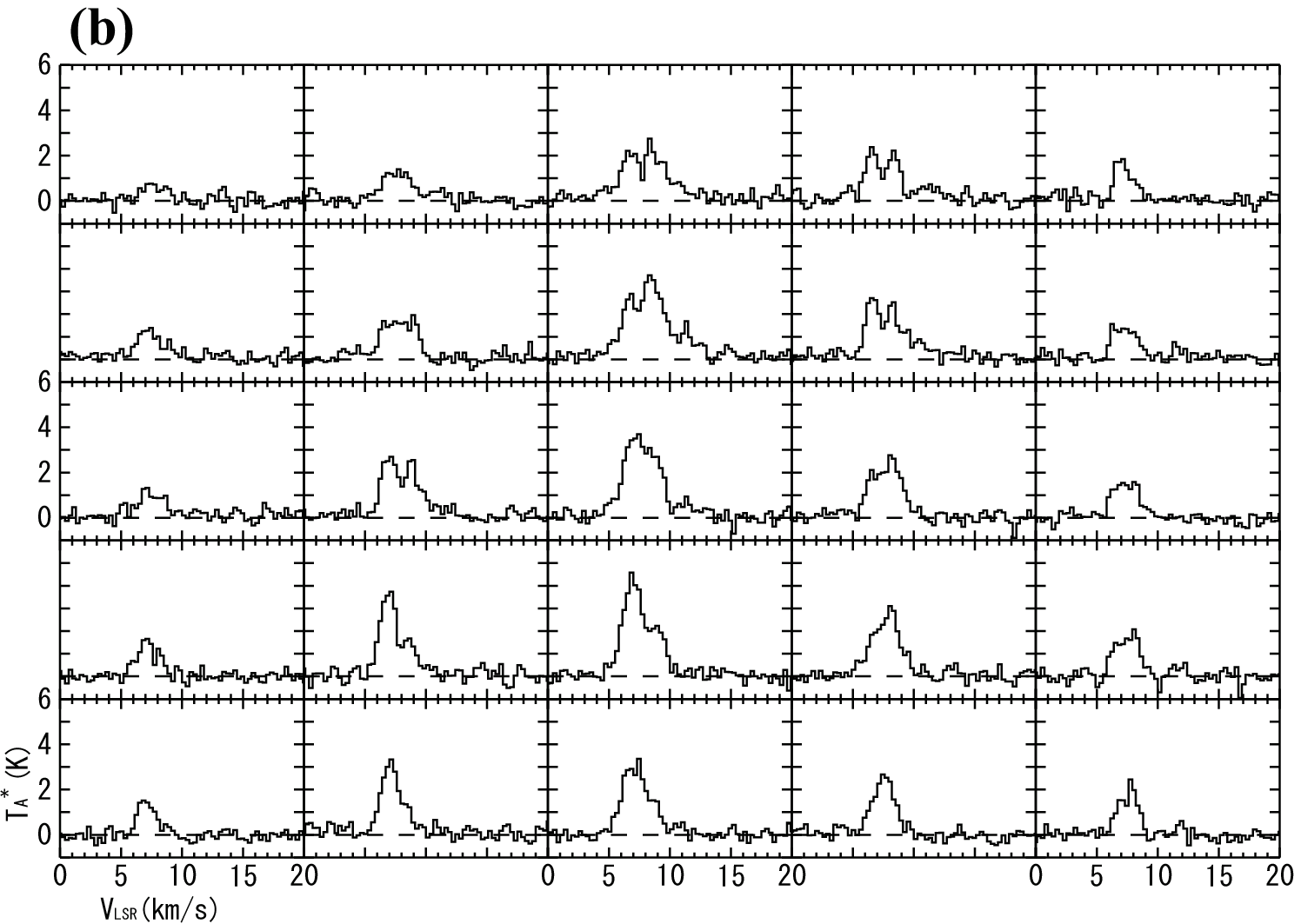}
\caption{{\it upper panel}: HCO$^+$ ($J=4-3$) total integrated intensity map 
 in the velocity range from $v_{\rm LSR} = 2.25$ km s$^{-1}$ 
to +15.25 km s$^{-1}$ toward the Serpens South protocluster.
The HCO$^+$ ($J=4-3$) integrated intensity is shown 
in color in units of K km s$^{-1}$.
The contours start at 2 K km s$^{-1}$ at an interval of 2 K km s$^{-1}$.
{\it lower panel}: 
HCO$^+$ ($J=4-3$) spectra. Each profile is averaged inside each 
box indicated in Figure \ref{fig:profile3}.}  
\label{fig:profile3}
\end{figure}

\begin{figure}
\epsscale{0.8}
\plotone{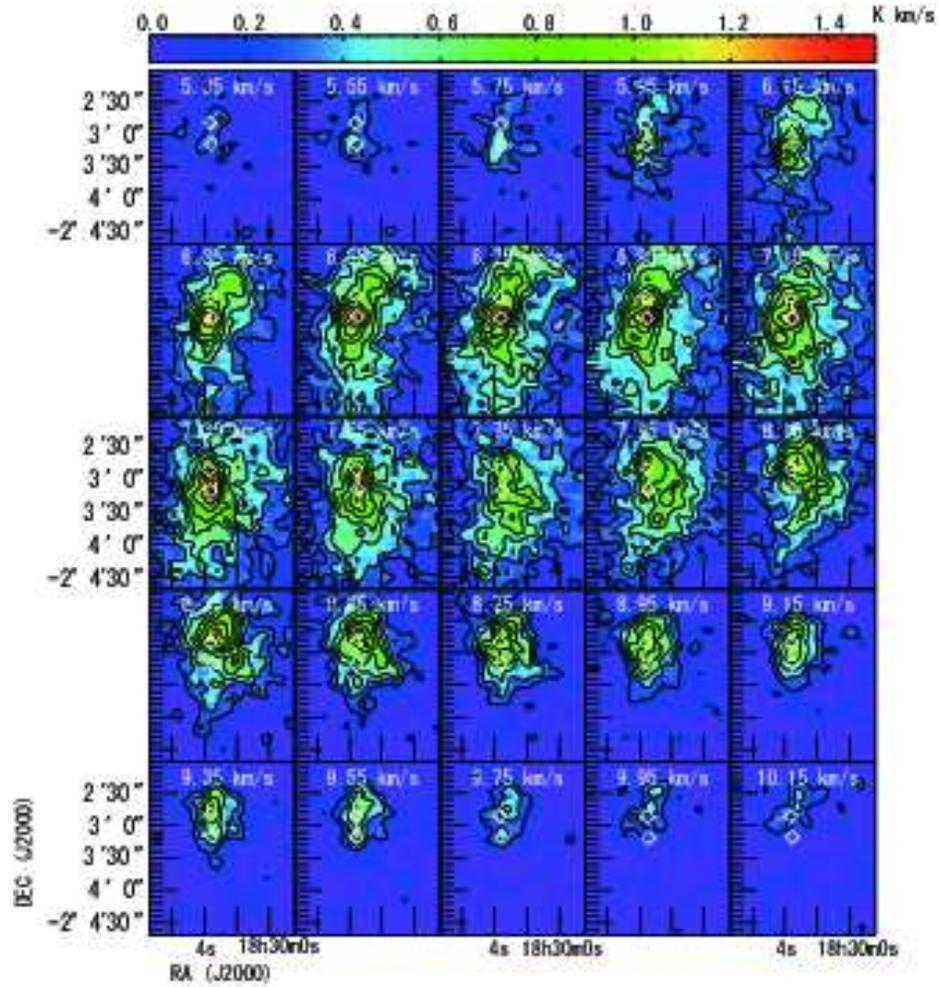}
\caption{
Velocity channel maps of the HCO$^+$ ($J=4-3$) emission
with the velocity intervals of 0.2 km s$^{-1}$.
The LSR velocity is indicated at the top of each panel.
The contour levels go up in 0.15 K km s$^{-1}$ step, 
starting from 0.2 K km s$^{-1}$.
The HCO$^+$ ($J=4-3$) channel maps indicate that the 
dense clump has clumpy structure. 
The positions of the possible driving sources of the outflows
are indicated by the white diamonds 
(sources 1 and 2 in Fig. \ref{fig:spitzer2}b).
Sources 1 and 2 are the possible driving sources 
of the R1 and R3 (and R2) outflow lobes, respectively.
The latter coincides with the position of the 1.1 mm peak 
(see Section \ref{subsec:outflow} in detail).}  
\label{fig:hcochannelmap}
\end{figure}

\begin{figure}
\epsscale{0.4}
\plotone{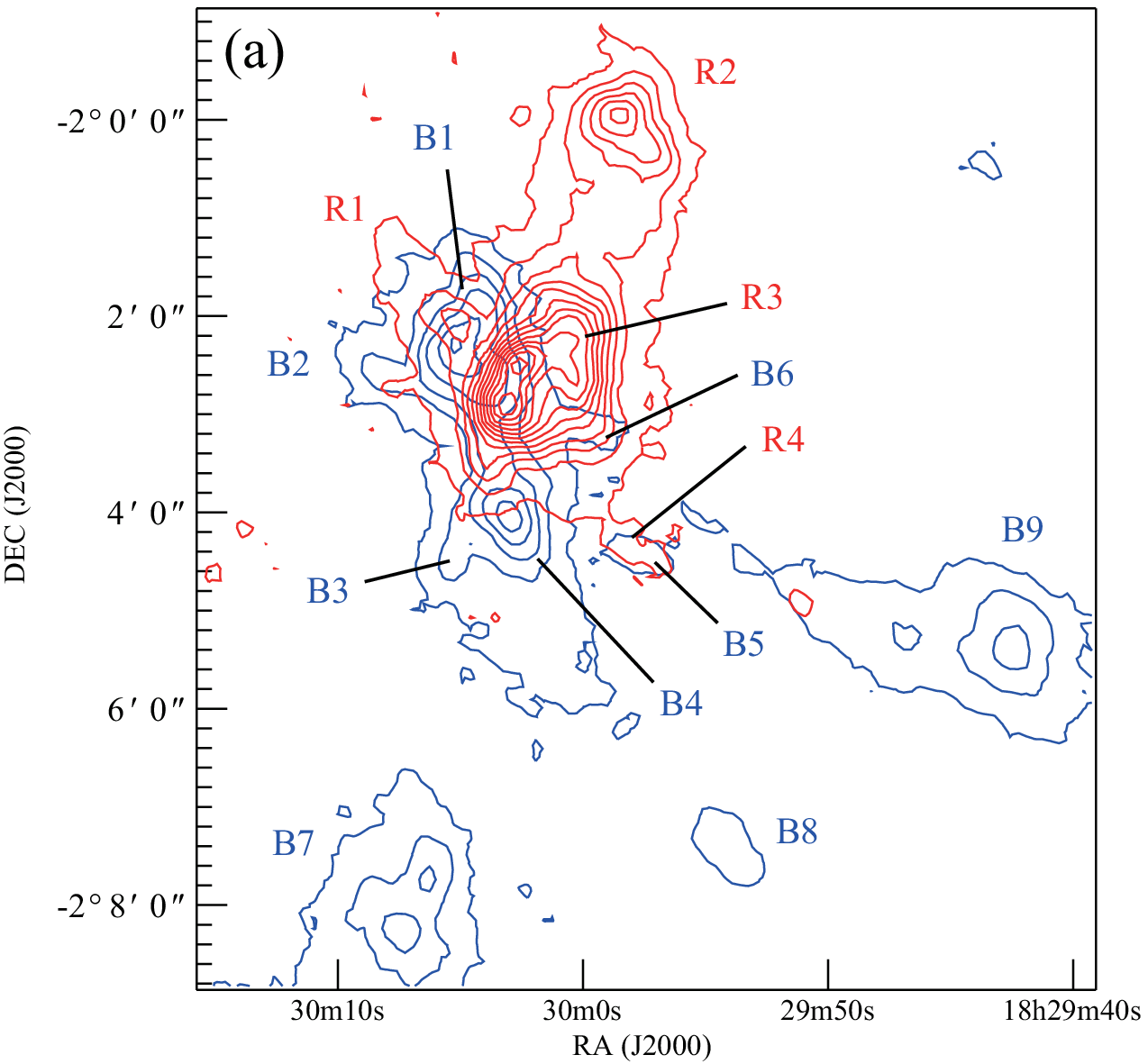}
\plotone{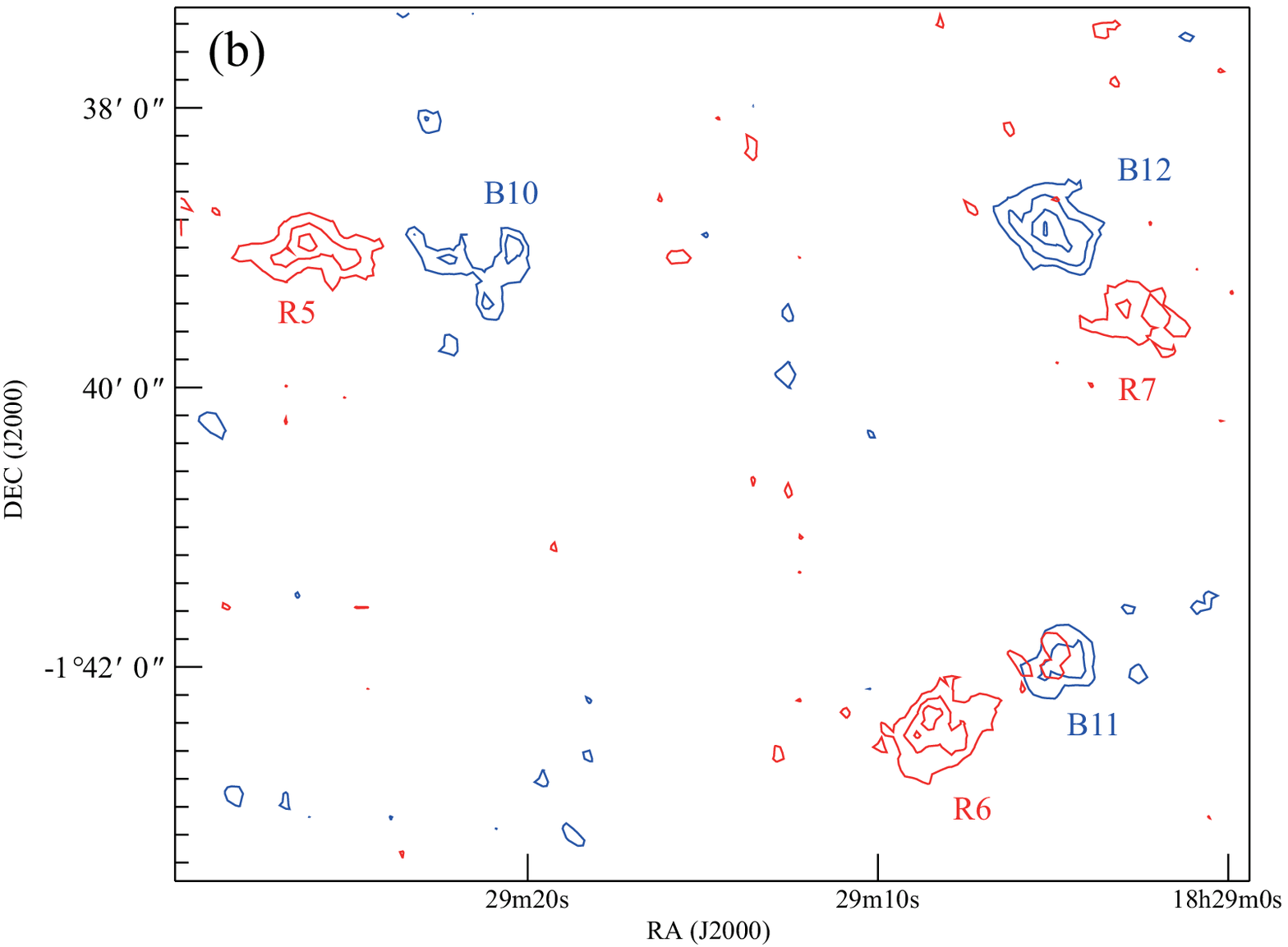}
\plotone{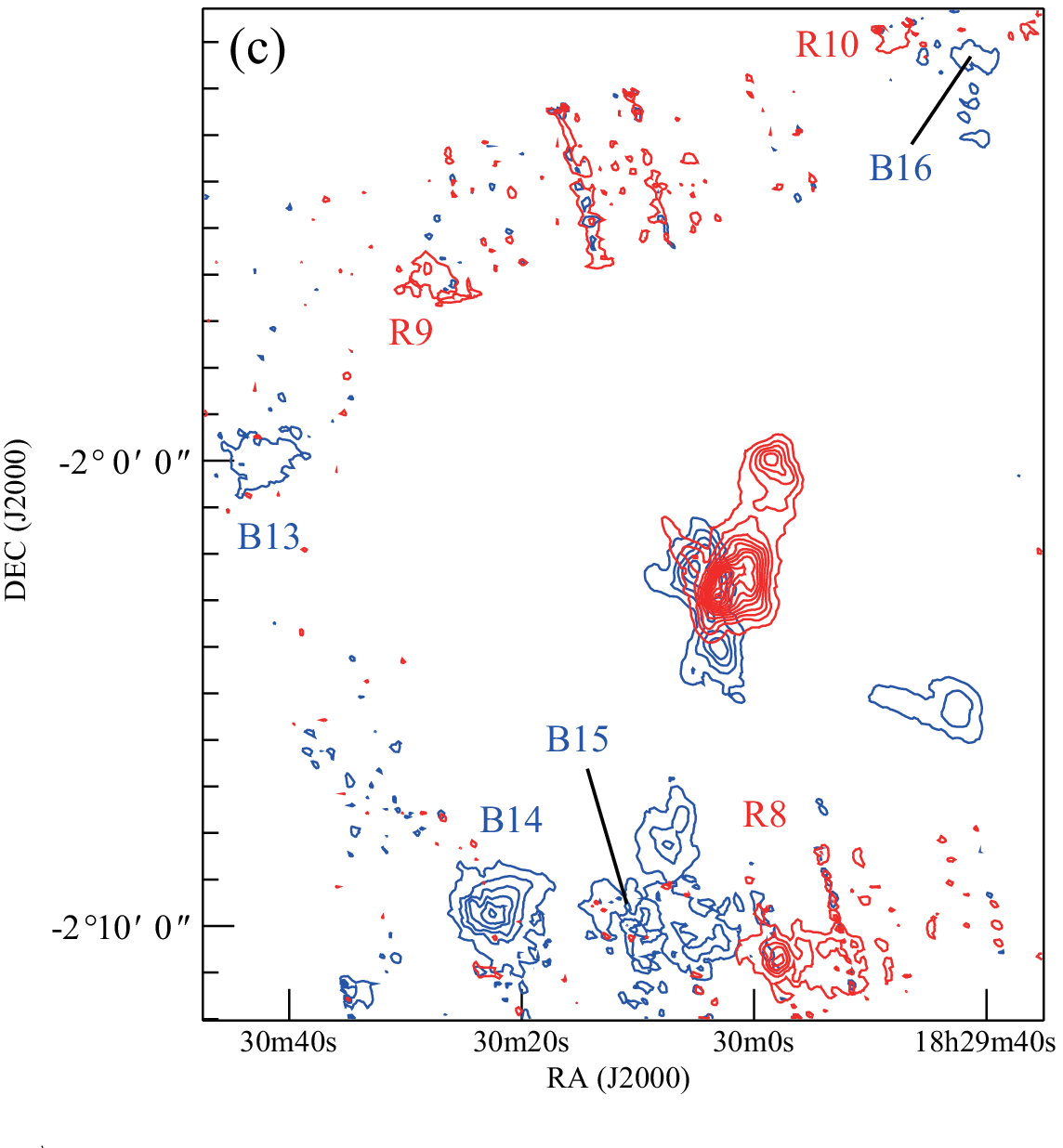}
\caption{
(a) Molecular outflow lobes identified from CO ($J=3-2$) emission
toward the Serpens South protocluster (Box 1).
The blue contours represent blueshifted $^{12}$CO gas and
red contours represent redshifted $^{12}$CO gas. The blue and red 
contour levels go up in 6 K km s$^{-1}$ step, 
starting from 3 K km s$^{-1}$.
The integration ranges are $-9.75$ to 3.75 km s $^{-1}$ for 
blueshifted gas and 11.25 to 29.25 km s$^{-1}$ for redshifted gas.
(b) Same as panel (a) but for the northern part of the Serpens 
South filament (Box 3).
The blue and red contour levels go up in 6 K km s$^{-1}$ step, 
starting from 6 K km s$^{-1}$.
(c) Same as panel (a) but for the southern part of the Serpens 
South filament (Box 2).
The blue and red contour levels go up in 6 K km s$^{-1}$ step, 
starting from 6 K km s$^{-1}$.
}  
\label{fig:bluered}
\end{figure}

\begin{figure}
\epsscale{1.0}
\plottwo{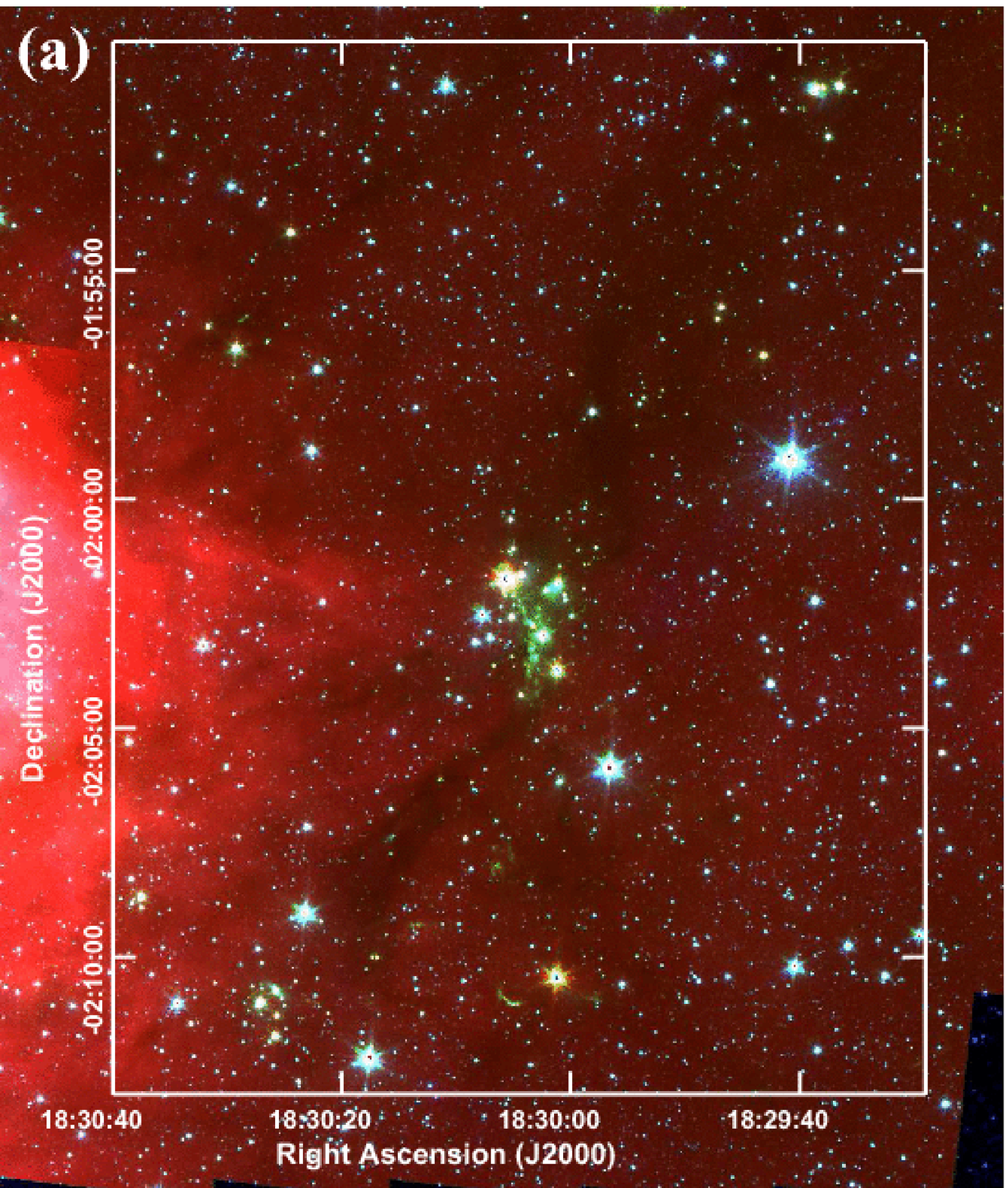}{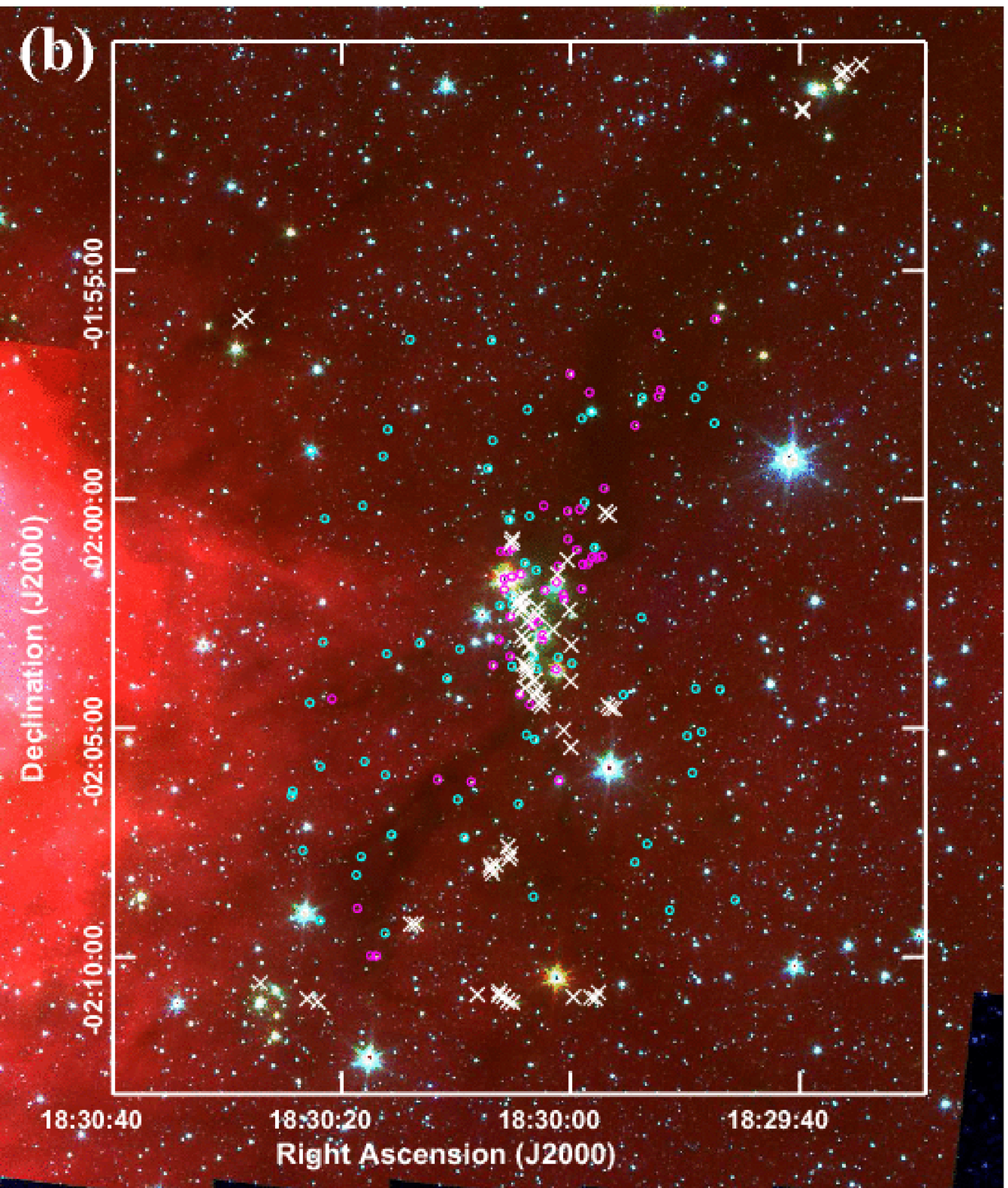}
\caption{(a) Three-color image of Serpens South with IRAC
band 1 (3.6 $\mu$m, blue), IRAC band 2 (4.5 $\mu$mm green), and IRAC
 band 4 (8.0 $\mu$m, red). 
(b) Same as panel (a) but the positions of 
Class I ({\it magenta circle}) and II ({\it cyan circle}) sources
identified by \citet{gutermuth08} are overlaid. 
The positions of the infrared H-H objects are indicated by the crosses.
}  
\label{fig:spitzer}
\end{figure}

\begin{figure}
\epsscale{1.0}
\plottwo{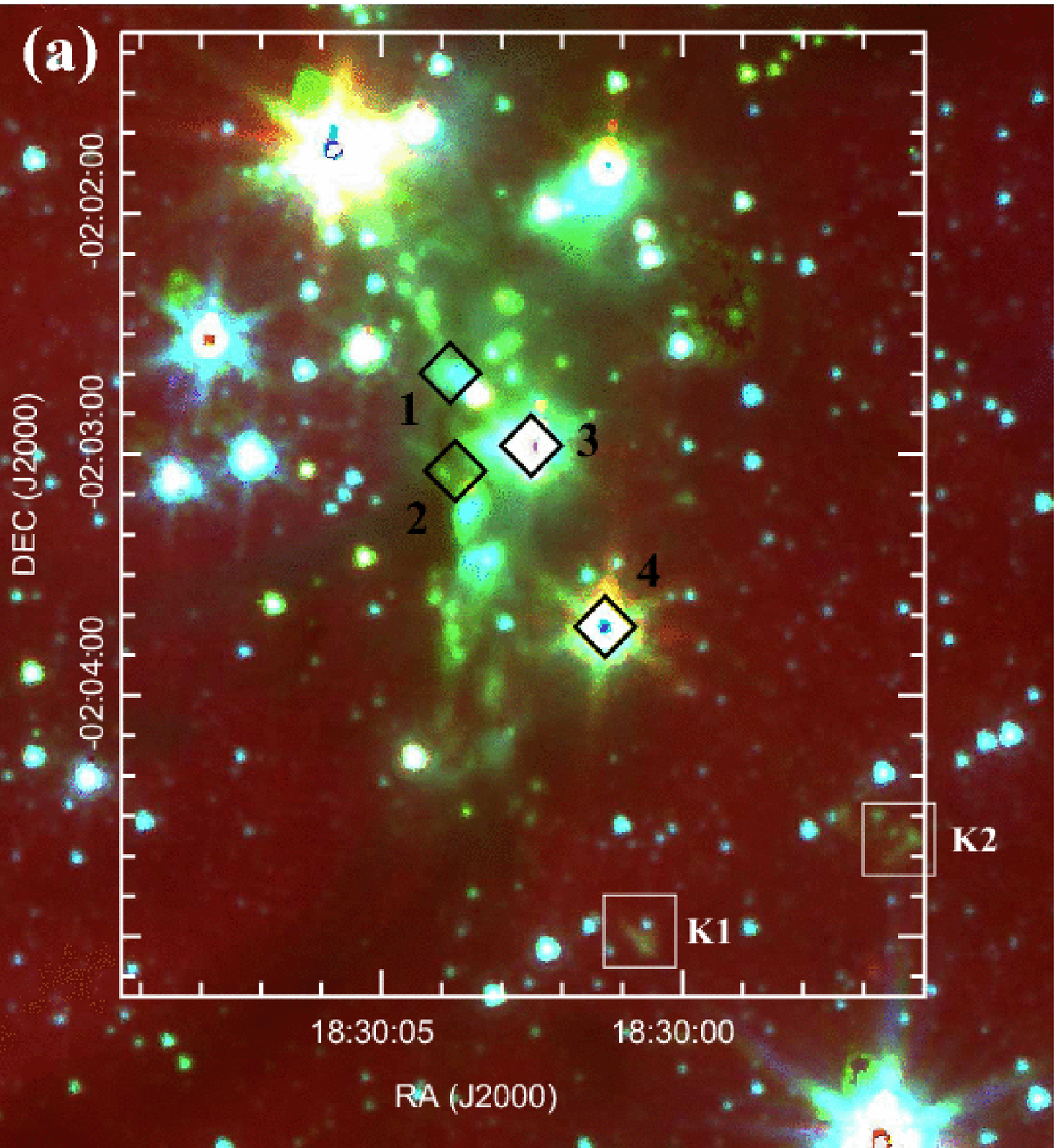}{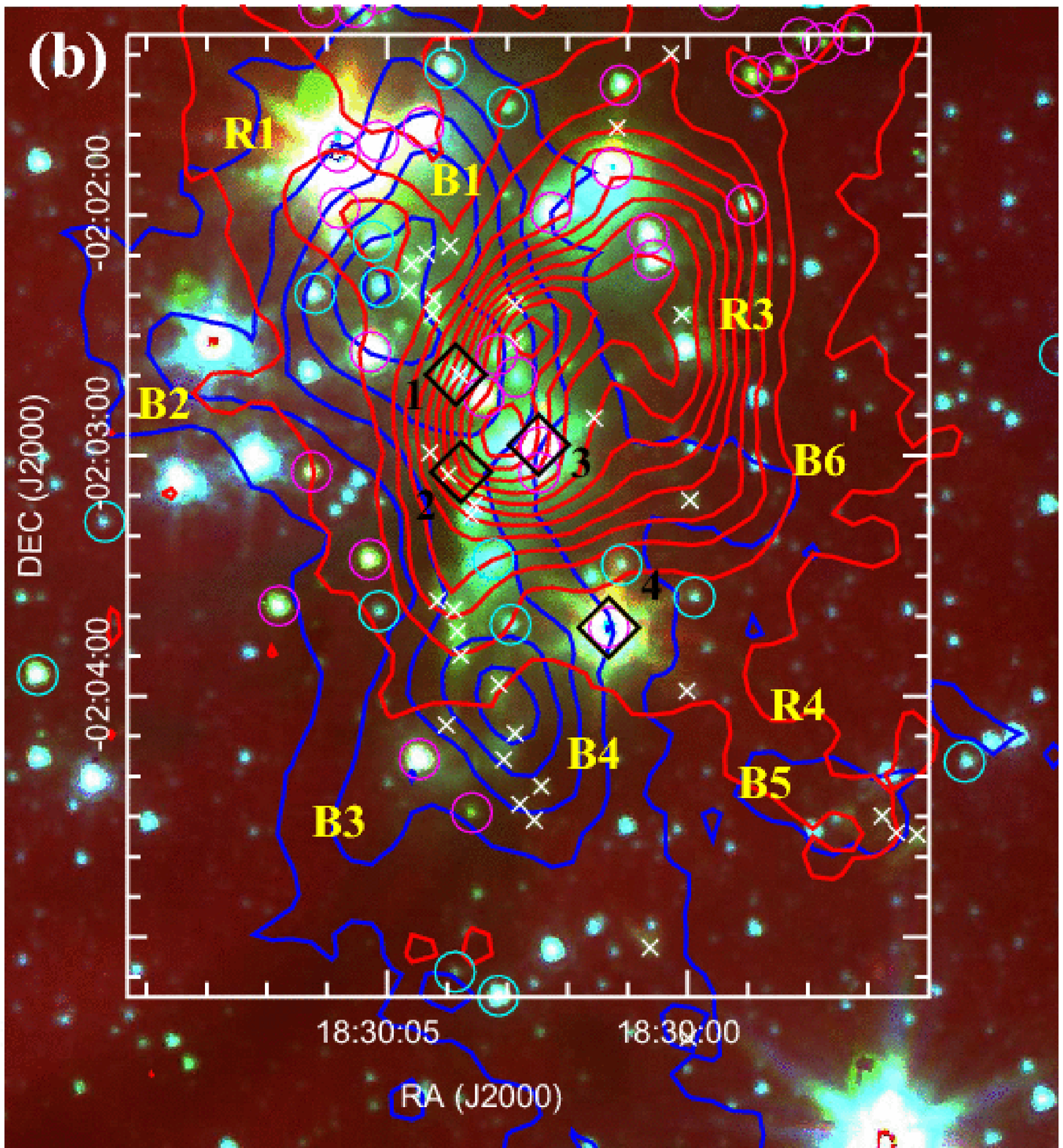}
\caption{(a) Blow-up of the Spitzer IRAC image toward the central region 
of Serpens South. 
The diamonds indicate the positions of the possible outflow 
driving sources (see Section \ref{subsec:outflow} in detail).
(b) Same as panel (a) but the blueshifted and redshifted 
CO ($J=3-2$) outflow lobes are overlaid on the image by the blue and red 
contours.  The contours and circles are the same as those of 
Fig. \ref{fig:bluered}a and Fig. \ref{fig:spitzer}, respectively.
}  
\label{fig:spitzer2}
\end{figure}

\begin{figure}
\epsscale{0.8}
\plotone{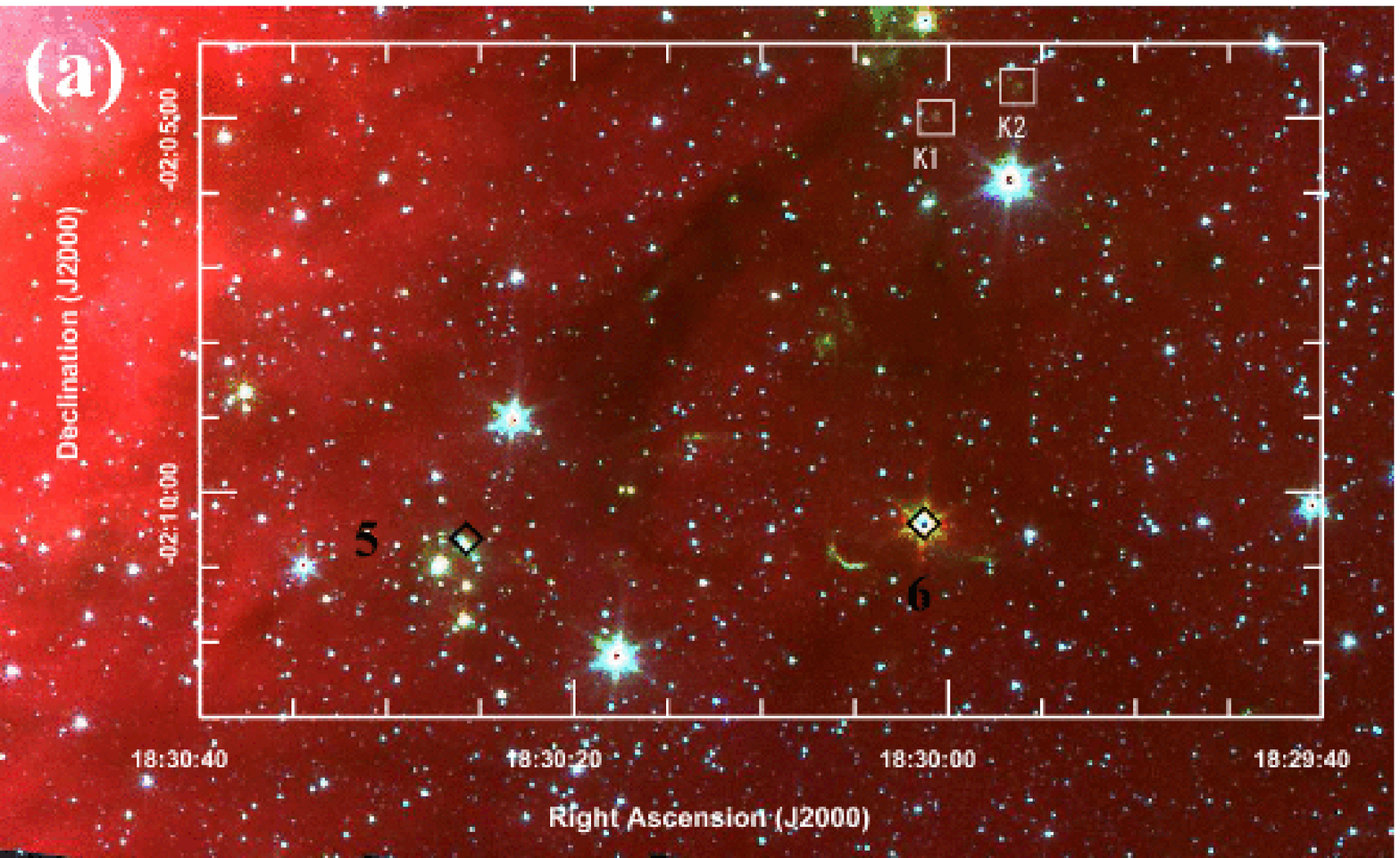}
\plotone{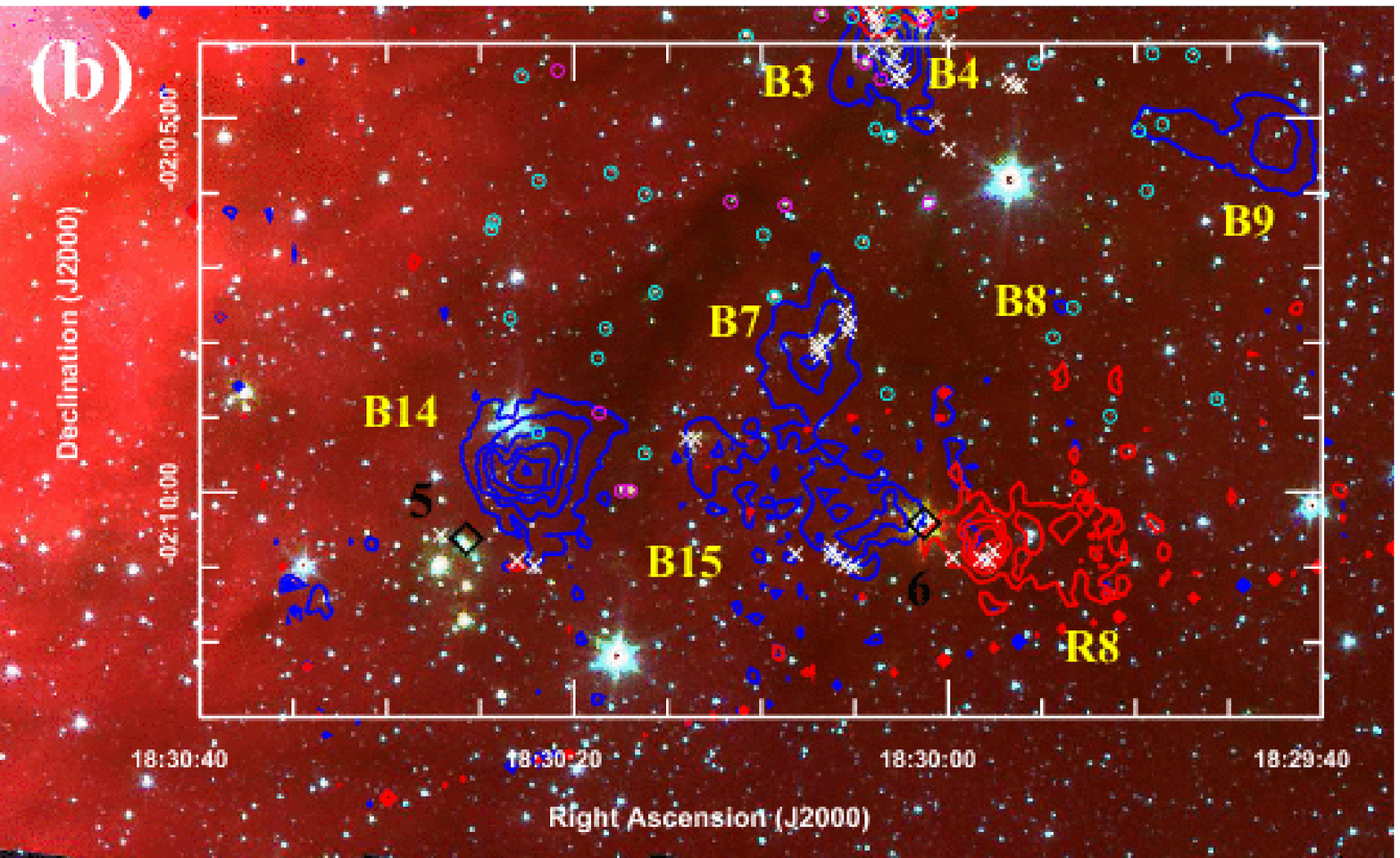}
\caption{(a) Blow-up of the Spitzer IRAC image toward the southern part
of Serpens South.
The squares indicate the positions of the same objects shown in 
Fig. \ref{fig:bluered3}a. 
The blow-ups of the areas labeled by K1 and K2 are presented in 
Fig. \ref{fig:spitzer4}.
(b) Same as panel (a) but the blueshifted and redshifted 
CO ($J=3-2$) outflow lobes are overlaid on the image by the blue and red 
contours.  The circles and crosses are the same as those of Fig. 
\ref{fig:spitzer}.
}  
\label{fig:spitzer3}
\end{figure}

\begin{figure}
\epsscale{1.0}
\plottwo{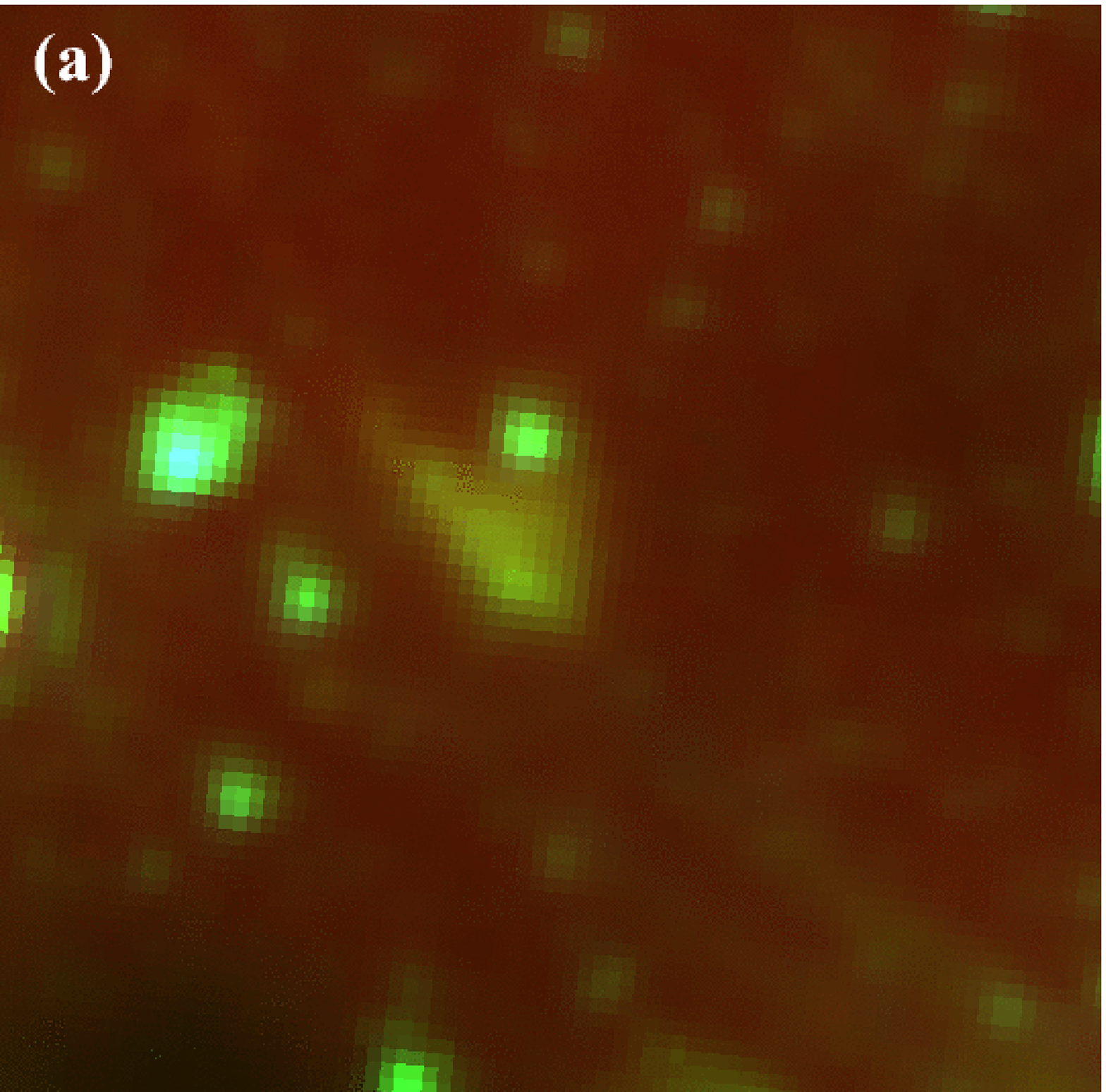}{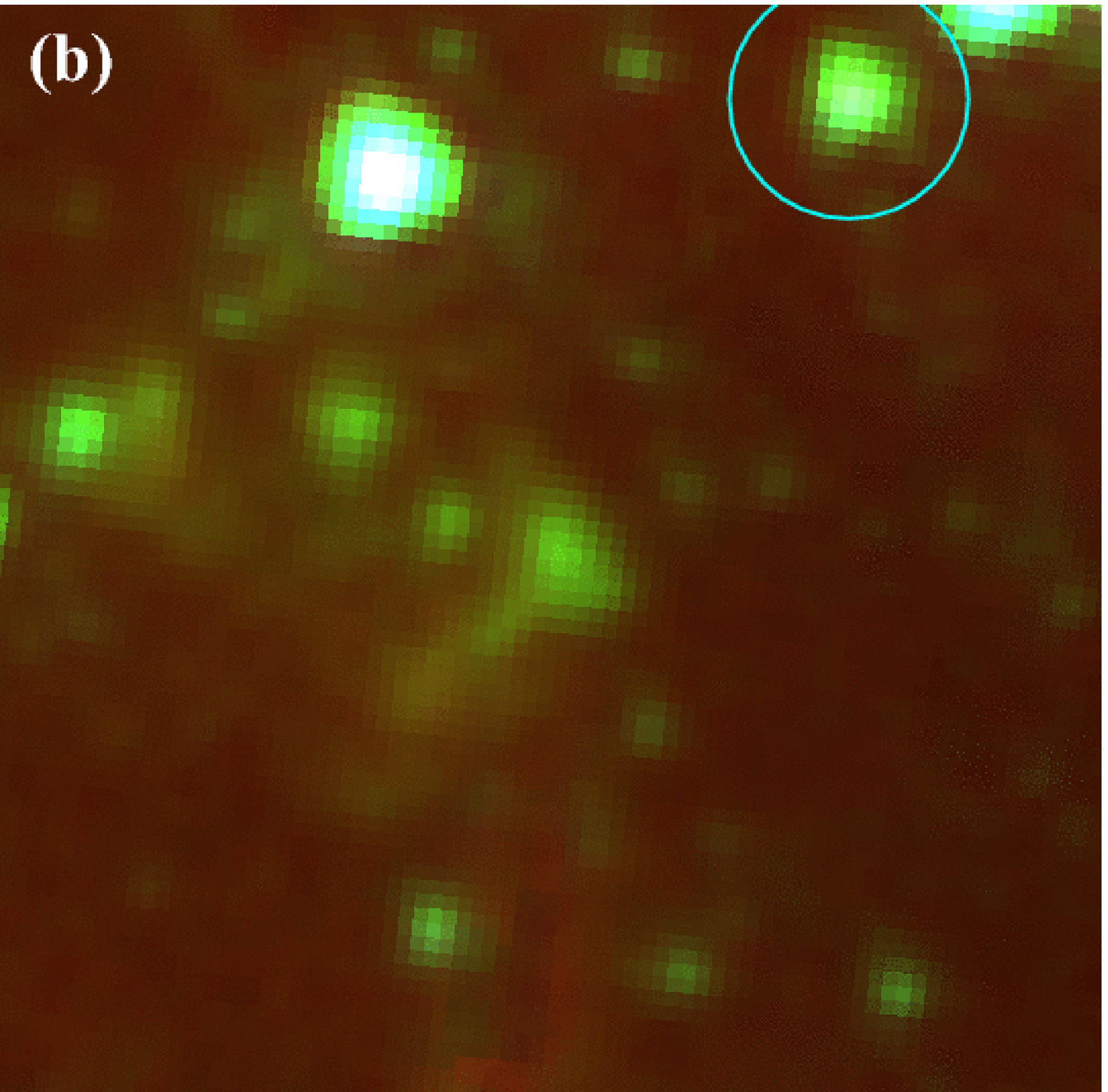}
\caption{Blow-ups of the areas indicated by K1 and K2 
in Figs. \ref{fig:spitzer2}a and \ref{fig:spitzer3}a.
An infrared H-H object having a bow shape
is seen near the center of each panel.
}  
\label{fig:spitzer4}
\end{figure}

Here, using the CO ($J=3-2$) data cubes, 
we attempt to identify high-velocity components that originate
from molecular outflows.
First, we scrutinize the velocity channel maps 
(Figs. \ref{fig:channelmap} and \ref{fig:channelmap2})
and the position-velocity diagrams (see e.g., Fig. \ref{fig:pvmap}) 
to find localized blueshifted or redshifted emission. 
As can be seen in Figs. \ref{fig:channelmap} and \ref{fig:channelmap2}, 
the $^{12}$CO high velocity structure is very complex and crowded
in the southern box with about $20' \times 20'$ area (Box 2). 
Thus, identifying outflows from the $^{12}$CO emission alone is 
extremely difficult in this crowded area.  
Therefore, we compare the CO image with three-color Spitzer IRAC images 
(3.6 $\mu$m in blue, 4.5 $\mu$m in green, 
8.0 $\mu$m in red) of this region, which are presented
in Figure \ref{fig:spitzer}.
The blow-up of the central area is shown in Figure \ref{fig:spitzer2}.
According to previous studies \citep[e.g.,][]{noriega-crespo04,takami10}, 
extended objects that have strong emission in IRAC channel 2 
(4.5 $\mu$m) compared to IRAC channel 1 (3.6 $\mu$m) and 
stand out as extended green objects in the three-color 
images are likely to be the objects shocked by the protostellar 
outflows. Here, we refer to such objects as infrared H-H objects.
To identify the infrared H-H objects, we first made an image of 
the remaining emission in the Spitzer IRAC band 2 (4.5 $\mu$m) image 
after the subtraction of the emission in band 1 (3.6 $\mu$m).
This image processing in practice allows us to discriminate between
shocks and stars, displaying them with positive and negative values,
respectively \citep{zhang09,takami10}. 
We then visually inspected the image to search for mid-infrared 
outflows. We selected spatially-extended sources that have significant 
excess at 4.5 $\mu$m as mid-infrared outflows.
The positions of the identified infrared H-H objects 
are indicated by the crosses in Figs. \ref{fig:spitzer}b, 
\ref{fig:spitzer2}b, and \ref{fig:spitzer3}b.
Some of the infrared H-H objects have bow-shapes.
For example, the infrared H-H objects labeled as K1 and K2 
in Fig. \ref{fig:spitzer3}a have arc-like features, 
both of which appear to move toward the south west direction 
(see Figs. \ref{fig:spitzer4}a and \ref{fig:spitzer3}b).
From our CO ($J=3-2$) and Spitzer IRAC data, 
we identified 15 blueshifted and 10 redshifted components.
The positions of the identified lobes are indicated 
in Fig. \ref{fig:bluered}.
Table \ref{tab:outflow lobe} gives a brief summary of the identified
outflow lobes.
We also show in Fig. \ref{fig:bluered2} the blueshifted and redshifted 
high-velocity components identified from HCO$^+$ ($J=4-3$).
In the following, we describe some of the main features 
of the high-velocity components we identified from the CO ($J=3-2$)
data.

\begin{figure}
\epsscale{1.0}
\plotone{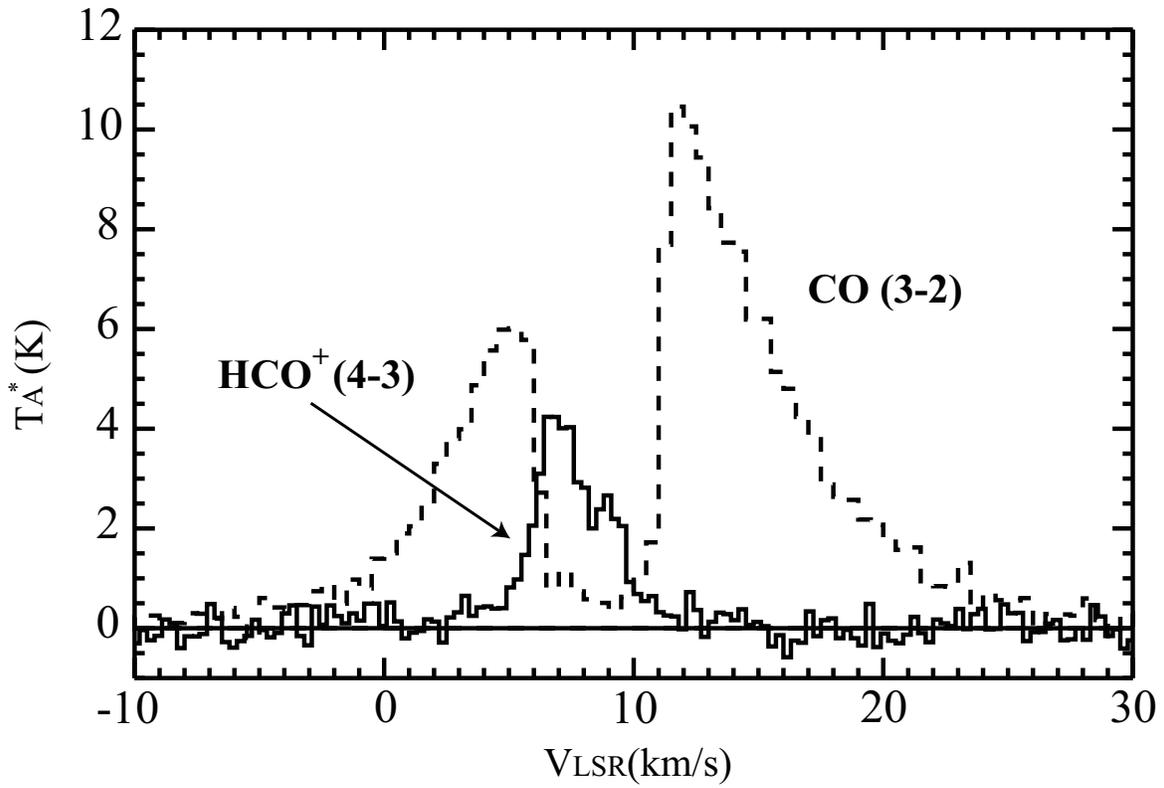}
\caption{HCO$^+$ ($J=4-3$) and CO ($J=3-2$) line profiles toward
source 2 indicated in Fig. \ref{fig:spitzer2}b.
The HCO$^+$ ($J=4-3$) shows a clear blue-skewed profile, indicating that 
the envelope gas is infalling toward the central object.
}  
\label{fig:blueskewed}
\end{figure}

The relatively strong blueshifted components B1 and B4 are well 
localized in the channel maps and appear at the maps within  
the wide range of 0.75 km s$^{-1}$ to 5.75 km s$^{-1}$.
These lobes have extremely high-velocity blueshifted components.
In particular, the B1 component has moderately-strong emission 
even in the velocity range of about $-$10 km s$^{-1}$ 
(see Fig. \ref{fig:profile1}).
There are also many infrared H-H objects associated with these two components. 
The B1 and B4 appear to move away from the central dense part
toward north and south, respectively.
In fact, the most distant infrared H-H object associated with B4 
has a bow shape, consistent with the interpretation 
that it is moving away from the dense part 
(see Fig. \ref{fig:spitzer2}b).
As mentioned below, the B4 component might originate from a compact 
3 mm source located at the 1.1 mm peak, recently identified by 
the Nobeyama Millimeter Array (NMA) observations (Fukuda et al. in prep.).
If this is the case, the corresponding redshifted components
are likely to be R2 and R3. The position of the 1.1 mm peak 
is indicated by a diamond (source 2 in Fig. \ref{fig:spitzer2}b) 
in the Spitzer image.  It appears
that no clear Spitzer sources are associated with the 1.1 mm peak.
The 3 mm source discovered by the NMA observations may be a 
deeply-embedded, extremely-young protostar
such as a Class 0 source.  In fact, the HCO$^+$ ($J=4-3$) line profile
toward source 2 shows blue-shifted part that are stronger 
than the red-shifted part (see Fig. \ref{fig:blueskewed}). 
Such a blue-skewed profile of an optically-thick line may be 
indicative of infalling gas, suggesting that the 3 mm source
located at the 1.1 mm peak is an infalling object having a powerful
outflow.

\begin{figure}
\epsscale{0.5}
\plotone{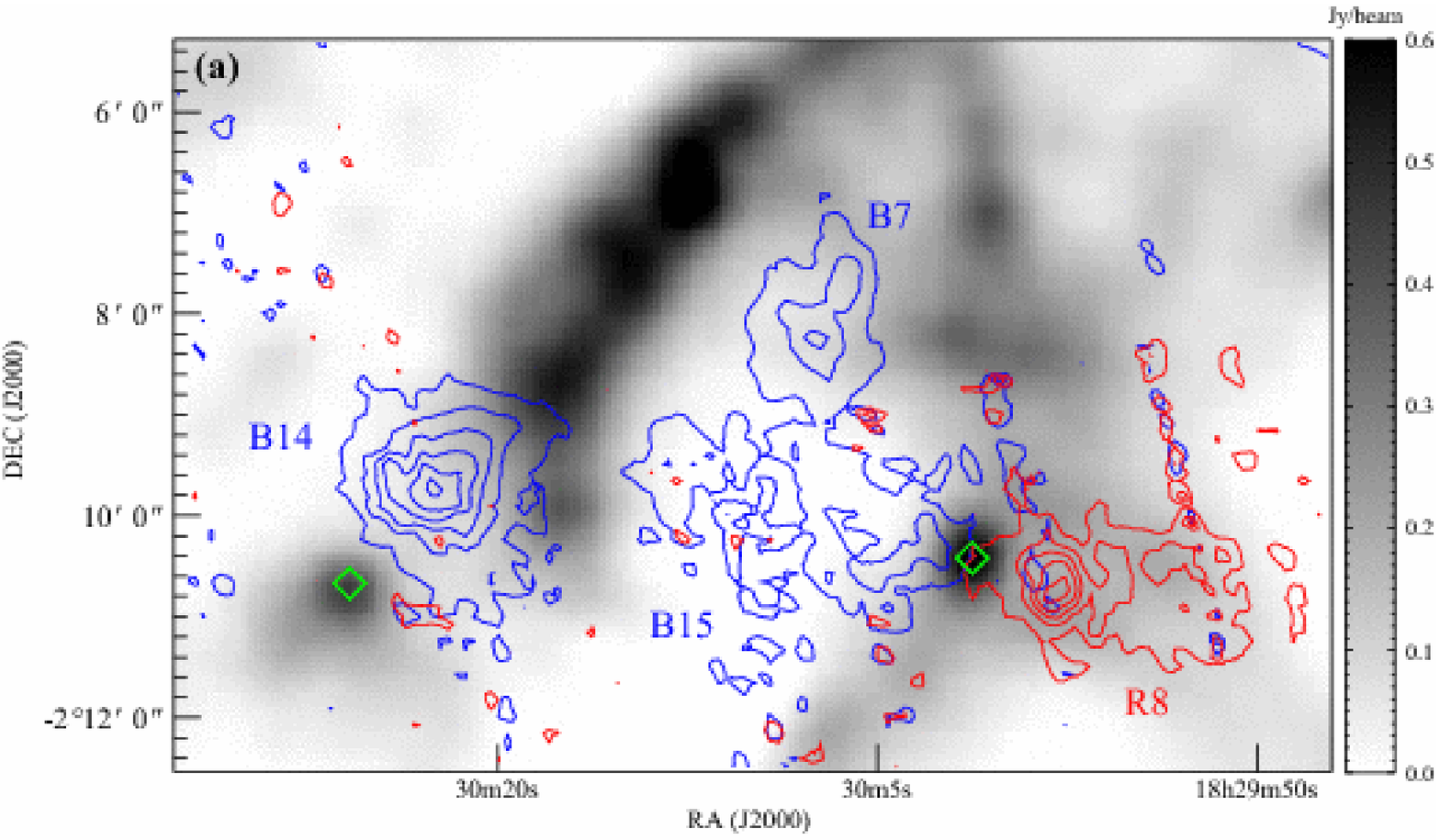}
\plotone{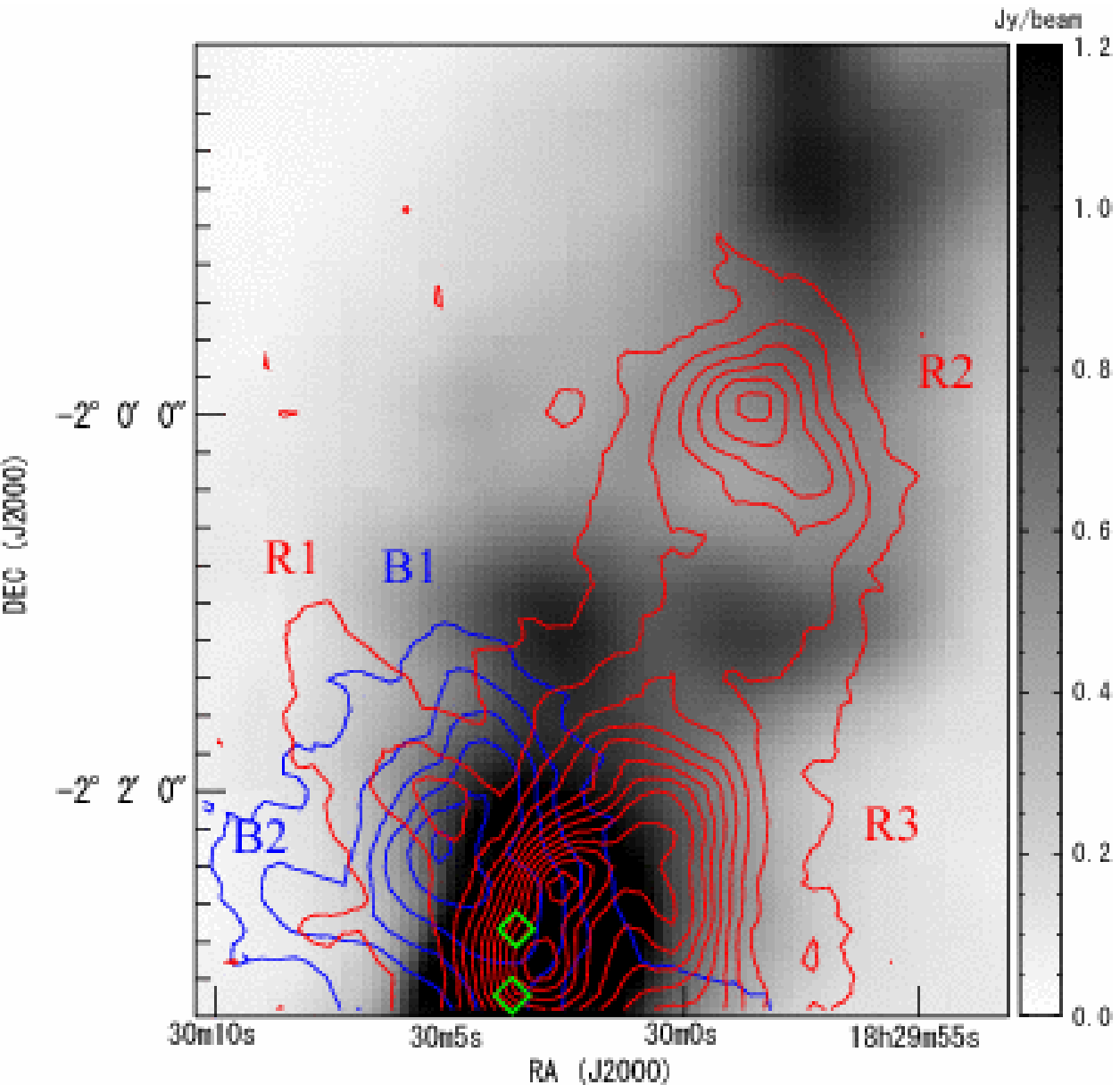}
\caption{
(a) Blow-up of the CO ($J=3-2$) outflow lobes located at the 
southern part of Box 2. The gray colors indicate 
the 1.1 mm continuum image. 
The upper and lower diamonds are the possible driving sources 
of R1 and R3 (and R2), respectively.
The latter coincides with the position of the 1.1 mm peak. 
(b) Same as panel (a) but for the northern part of Box 1.
The possible driving sources of the outflows
are indicated by the green circles whose positions are determined from
the positions of the sources seen in the Spitzer IRAC image.
In panels (a) and (b), the blue contours, red contours, and 
labels are the same as those of  Fig. \ref{fig:bluered}. 
}  
\label{fig:bluered3}
\end{figure}

\begin{figure}
\epsscale{1.0}
\plotone{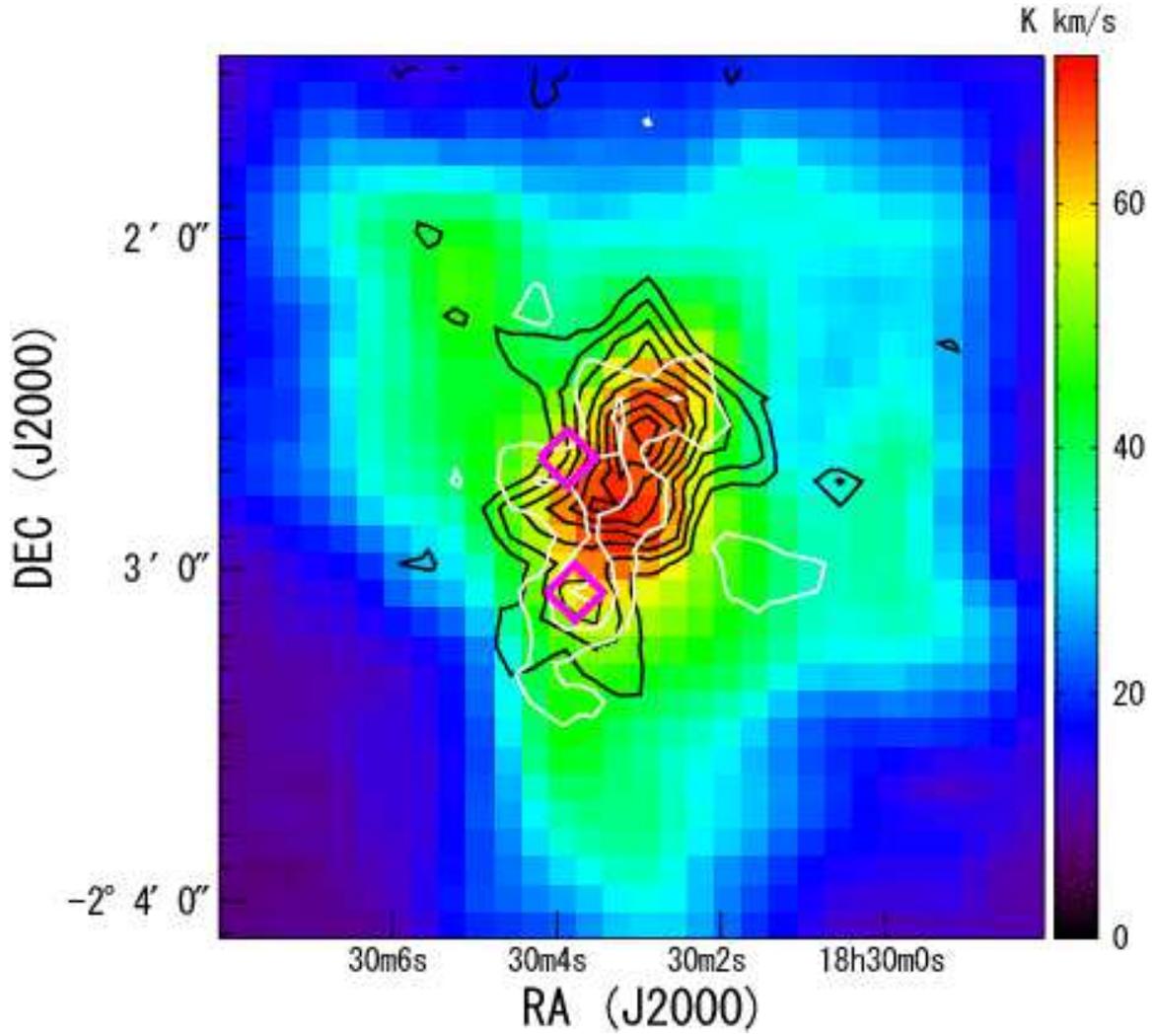}
\caption{
High velocity components identified from HCO$^+$ ($J=4-3$) emission
toward the Serpens South protocluster.
The color image indicates the CO ($J=3-2$) integrated intensity map
in the range of 2 km s$^{-1}$ to 15 km s$^{-1}$.
The white and black contours represent blueshifted and 
redshifted gas detected by HCO$^+$ ($J=4-3$), respectively. 
The white and black contour levels go up in 0.5 K km s$^{-1}$ and 
1.0 K km s$^{-1}$ steps, 
starting from 1.0 K km s$^{-1}$ and 1.0 K km s$^-1$, respectively.
The integration ranges are $3.25$ to 5.75 km s $^{-1}$ for the 
blueshifted gas and 9.75 to 15.25 km s$^{-1}$ for the redshifted gas.
The squares indicate the positions of the same objects shown in 
Fig. \ref{fig:bluered3}b.
}  
\label{fig:bluered2}
\end{figure}

\begin{figure}
\epsscale{1.0}
\plotone{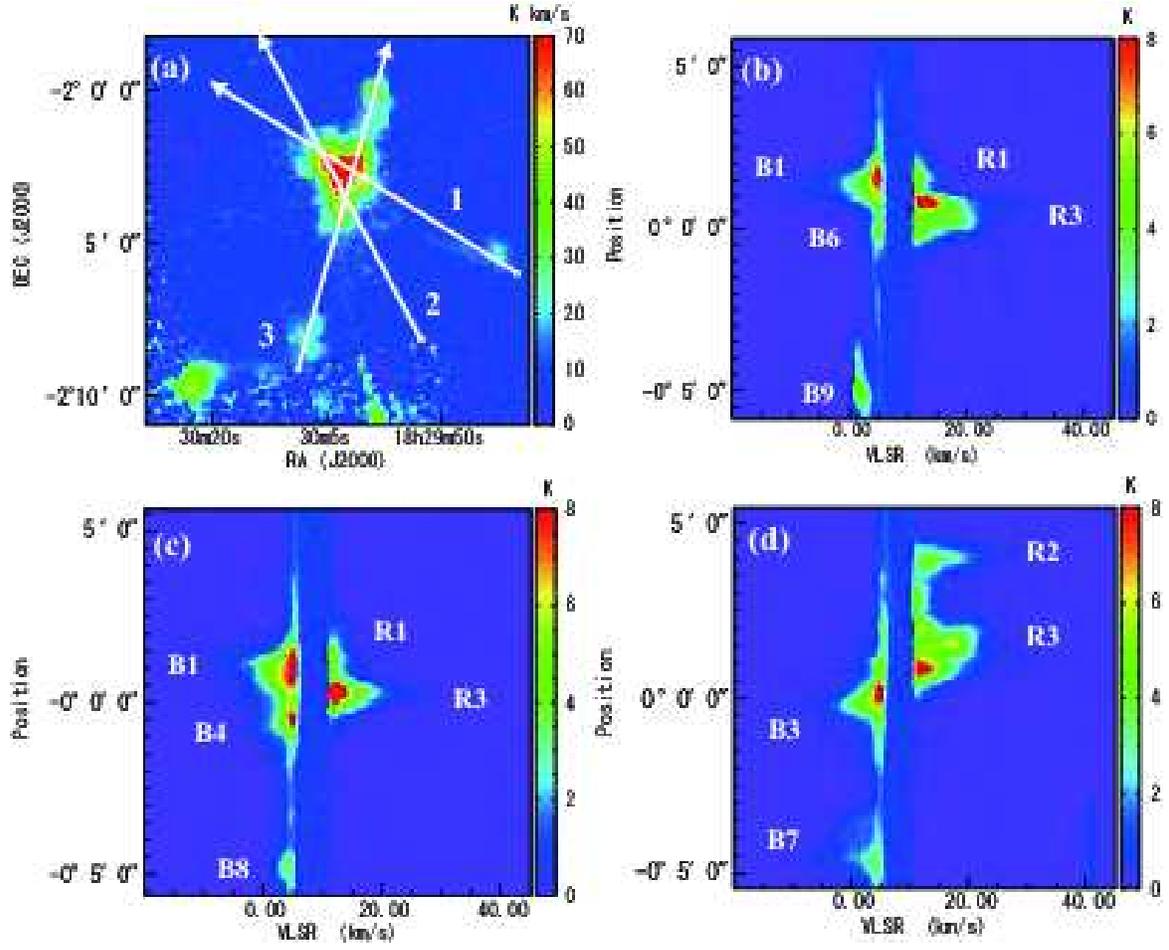}
\caption{(a) Integrated intensity map showing the positions of the PV
 diagrams. (b) Position-velocity diagram along the line 1 indicated 
in panel (a). (c) the same as that of panel (b) but for the line 2, 
(d) the same as that of panel (b) but for the line 3. 
}  
\label{fig:pvmap}
\end{figure}

There are two faint blueshifted components labeled with
B2 and B3. They are vaguely seen in the channel maps with high velocities
(0.75 km s$^{-1}$ $\sim$ 1.75 km s$^{-1}$). Some infrared H-H objects appear
to be associated with the B3 component (see Fig. \ref{fig:spitzer2}).
The position-velocity diagram indicates that the blueshifted 
component B8 is likely to be the outflow component that appears
to move away toward south.
The blueshifted components B14 and B7 are located near the edge of Box 1.
Although B7 appears to move from north west to south east from 
its morphology, the driving source remains unknown.
There is no redshifted counterpart associated with B7.  
The B7 component is also associated with some extended 
green objects suggesting that the ambient gas is shocked 
by the B7 component (see Fig. \ref{fig:spitzer3}). 
The B14 component also do not have a redshifted counterpart.
There is a compact dust continuum source immediately south of B14, 
where \citet{bontemps10} found two protostar candidates by the
Herschel observations.  The Spitzer image also shows two 
compact sources at the same places as the dust continuum source
and the Herschel protostar candidates (Fig. \ref{fig:spitzer3}).
We suggest that the northern source is likely to be the driving source
of B14. This interpretation appears to be supported by the fact that
the 1.1 mm emission is weak at the position of the B14 component, 
apparently breaking the main filament into two parts 
(see Fig. \ref{fig:bluered3}a), and bending the northern part of the
filament toward south.  
There are some infrared H-H objects in the southern part of B14.
These objects may originate from B14.

The relatively-faint blueshifted component B6 is clearly seen in the 
channel maps in the range of 3 km s$^{-1}$ to 5 km s$^{-1}$, 
apparently moving toward the south-west direction.
The possible driving source is one of the Class I sources identified by 
\citet{gutermuth08} (source 3 in Fig. \ref{fig:spitzer2}) because 
some H-H objects appear to be ejected from it.
The B6 component appears to point toward the relatively-strong 
blueshifted component B9.
However, it is unclear whether B9 originates from the outflow lobe. 
The peak of B9 is about 0.4 pc away from the 1.1 mm peak.
The position-velocity diagram may imply 
that this component has different LSR velocity 
($V_{\rm LSR} \sim 2$ km s$^{-1}$) from 
the central dense part and does not show 
the velocity structure typical of the outflow lobe
(see Fig. \ref{fig:pvmap}b).
However, there are no YSOs associated with the same area as the B9
component. 
Therefore, this component might be a fossil of the outflow lobe 
or infalling gas toward the central dense part along a sub-filament.
Further investigation is needed to clarify the origin of this component.

There is a redshifted component R1 between the two blueshifted
components B1 and B2. This redshifted component is clearly seen 
in the peak intensity map and appears to move away from the dense part.
This component appears to be associated with a blueshifted component, which 
can be seen in the velocity channel maps with 4.75 km s$^{-1}$ and 
5.75 km s$^{-1}$.  Therefore, this component may be almost parallel 
to the plane-of-sky direction.  The possible driving source
is indicated by a diamond in Fig. \ref{fig:spitzer2}a (source 1),
from which many H-H objects are apparently ejected.

The strong redshifted components R2 and R3 appear to move away 
from the central dense part, clearly showing the velocity structure 
typical of the outflow components. 
Several infrared H-H objects are associated with these components.
Interestingly, at the area between R2 and R3, the 1.1 mm emission 
shows a double-peak feature and the 1.1 mm emission tends to be weak
on the line connecting between R2 and R3
 (see Fig. \ref{fig:bluered3}b). In addition, 
there is an area having relatively strong 1.1 mm emission
just at the north of R2.
This morphology leads us to the following possibility:
the redshifted lobes R2 and R3 come from the same source and R2 went through 
the dense part, reaching and hitting the dense part located 
at the north. The R3 component appears to coincide with the 
high-velocity component identified by the HCO$^+$ ($J=4-3$) map
(see Fig. \ref{fig:bluered2}).
Such high-velocity dense gas may suggest that the outflow 
is very young.
As mentioned above, the possible driving source is 
a 3 mm source recently discovered by the NMA observations,
whose position almost coincides with the 1.1 mm peak
(source 2 in Fig. \ref{fig:spitzer2}, Fukuda et al. in prep.). 
However, other unknown sources might be the real driving sources
of these outflow lobes because this central part is very dense and 
crowded.

There is a faint redshifted component R4 in the south west part of the 
dense part.  Several infrared H-H objects are associated with the R4 component.
This component is clearly seen in the channel maps 
with 11.75 through 13.75 km s$^{-1}$, apparently 
moving toward the south-west direction. 
In fact, the most distant infrared H-H object 
(K2 in Fig. \ref{fig:spitzer4}b) has a bow shape, consistent with
the interpretation that it is moving away from the dense part.
Since a faint blueshifted component labeled by B5 is 
 associated with the R4 component,  
this component might be almost parallel to the plane-of-sky.
The possible driving source is one of the Class I objects
identified by \citet{gutermuth08} 
(source 4 in Fig. \ref{fig:spitzer2}) because
some infrared H-H objects are apparently ejected from it.

Several bipolar outflows can be identified clearly in the 
CO ($J=3-2$) data. For example, the southernmost outflow
with the B15 and R8 lobes looks relatively powerful and 
clearly shows a bipolar feature.  The driving source
is probably one of the Herschel candidate protostars
identified by \citet{bontemps10} (see their Fig. 5)
whose position is 
(RA, Dec) $\simeq$ (18h30m2s, $-2^\circ$10$'$20$''$). 
The candidate protostar is also seen in the 1.1 mm continuum 
image presented in Fig. \ref{fig:bluered3}b and 
the Spitzer IRAC image in Fig. \ref{fig:spitzer}, as a point-like source.
Another bipolar outflow can be seen 
near the upper right corner of panel (c) of Fig. \ref{fig:bluered}.
The driving source may be either a 
Herschel candidate protostar or a Class 0 object
located at the upper-right corner of Fig. 5 of \citet{bontemps10}.
In the northern box with about $13' \times 13'$ area (Box 3)
we found three molecular outflows, all of which 
show clear bipolar features: B10 and R5, B11 and R6, and 
B12 and R7.  Since this region is out of our Spitzer image,
we cannot identify the driving sources.

In Box 2, we identified four additional high-velocity components labeled by 
B13 and R9. These components possibly originate from 
some outflows. However, further observations would be needed to 
clearly identify the outflow lobes.

We note that the directions of the outflow lobes 
often appear to be roughly along the global magnetic field direction 
that is running almost perpendicularly to the main filament 
\citep{sugitani11}, e.g., B2, B5, B6, B9, B15, B16, R4, R8, and R10.  
However, several components, e.g., B1, B3, B4, R2, and R3, 
appear to be independent of the global magnetic field direction.  
This indicates that the central dense part is magnetically
supercritical and therefore, the local supersonic turbulence
have presumably perturbed the magnetic field significantly.
In fact, the recent numerical simulations of cluster formation
indicate that even the moderately-strong magnetic field 
causes a significant misalignment between the global field 
direction and the outflow propagation directions 
because of the strong turbulence \citep{nakamura11b}.

\subsection{Derivation of the Physical Quantities}

\subsubsection{Outflow Parameters}

Figure \ref{fig:bluered} shows that the high-velocity 
components are crowded in the central dense part where 
the protocluster resides and therefore  
it is very difficult to clearly discriminate the individual outflow 
lobes  from the data. 
Therefore, we postpone to accurately evaluate the physical 
properties of the individual outflows 
until high spatial resolution data taken by interferometric 
observations become available.  
In the following, we calculate some global physical properties 
of the outflowing gas located at the cluster center. 
On the other hand, five bipolar outflows mentioned in the previous 
subsection can be easily identified in the CO ($J=3-2$) map, 
because they are found in relative isolation.  
However, we postpone to calculate the individual outflow parameters 
of these bipolar outflows because the rms noise level
is relatively high with a significant scanning effect, 
precluding accurate estimation of the 
outflow parameters.

Examination of the spectra in the cloud shows that ambient $^{12}$CO 
spectra away from the
outflows appear to cover a velocity range of 6 to 11 km s$^{-1}$. 
We thus assume that the emission in this velocity range
include ambient emission for the entire cloud.
Therefore, the integration ranges used are $-15$ to 4 km s$^{-1}$
for the blue-shifted emission and 11 to 30km s$^{-1}$
for the red-shifted emission, respectively.
The outflow masses were calculated under the assumption of LTE, a
distance of 260 pc, and optically-thin emission in the line
wings with an excitation temperature of 30K. 
We note that the physical quantities estimated below are 
insensitive to the excitation temperature. 
The estimated quantities increase only by 20 \% over 
the range of $T_{\rm ex}= 20 - 50$ K.

Following \citet{nakamura11}, 
the outflow mass $M_{\rm out}$, 
the outflow momentum $P_{\rm out}$, 
and kinetic energy $E_{\rm out}$ are estimated as
\begin{equation}
M_{\rm out}=\sum_j M_j \ , 
\end{equation}
\begin{equation}
P_{\rm out}=\sum_j M_j |V_j - V_{\rm sys}| / \cos \xi  \ , 
\end{equation}
and
\begin{equation}
E_{\rm out} = \sum _j \frac{1}{2} M_j (V_j-V_{\rm sys})^2  / \cos^2 \xi \ ,
\end{equation}
respectively, where 
$M_j$ is the mass of the $j$-th channel, $V_j$ is the LSR velocity of the $j$-th channel and 
$V_{\rm sys}$ is the systemic velocity of the driving source.
Here, we adopt $V_{\rm sys}=7.5$ km s$^{-1}$.
The angle $\xi$ is the inclination angle of an outflow,
which is generally uncertain. Here, we adopt $\xi = 57.3^\circ$, 
following \citet{bontemps96}.
The mass of the $j$-th channel, $M_j$, is expressed as 
\begin{eqnarray}
M_{j} &=& \mu_g m_{\rm H_2} X_{\rm CO}^{-1} \Omega D^2
\sum _i N_{\rm CO,\it i,j} \nonumber \\
&=& 3.2\times 10^{-9} \left({X_{\rm CO} \over 10^{-4}}\right)^{-1}
\left({D \over 260 {\rm pc}}\right)^2
\left({\Delta \theta \over {\rm arcsec}}\right)^2
 \nonumber \\
&\times& \left({\eta \over 0.57}\right)^{-1}
\left({\sum _i T_{A,i,j}^* \Delta v \over 
{\rm K \ km \ s^{-1}}}\right)
T_{\rm ex} \exp\left[{33.2 \ {\rm K} \over T_{\rm ex}}\right] M_\odot
 \ ,
\end{eqnarray}
where $i$ is the grid index on the $j$-th channel, 
$N_{\rm CO, \it i,j}$ is the CO column density at the $i$-th
grid on the $j$-th channel, 
$\eta$ (=0.57) is the main beam efficiency of the ASTE telescope, 
$m_{\rm H_2}$ is the mass of a hydrogen molecule, 
$X_{\rm CO}$ is the fractional abundance of CO relative to 
H$_2$, and the mean atomic weight of the gas $\mu_g$ is set to 1.36.
We adopt $X_{\rm CO}=10^{-4}$.
$\Omega$ is the solid angle of the object and $D$ is the
distance to the object.
From these quantities, we calculate the characteristic velocity
as $V_{\rm out} = P_{\rm out}/M_{\rm out}$.

Table \ref{tab:outflow} summarizes the outflow parameters 
derived from the CO ($J=3-2$) data.
The physical quantities presented in Table \ref{tab:outflow}
depend on the adopted velocity interval for the integration.
For example, if we change the velocity intervals as
$-15$ to 6 km s$^{-1}$ ($-15$ to 2 km s$^{-1}$) 
for the blue-shifted emission and 
9 to 30km s$^{-1}$ (13 to 30km s$^{-1}$ ) 
for the red-shifted emission,
the total mass, momentum, and energy change by +65\% ($-$50\%), 
20\% ($-$30\%), and 5 \% ($-$13 \%), respectively.
This indicates that the total momentum and energy do not depend 
strongly on the adopted velocity intervals, although the total mass
does. This is probably due to the fact that the outflow components with lower
velocities have lower momenta and kinetic energies.
As mentioned above, the distance to Serpens South remains uncertain.
If we adopt $D=410$ pc \citep[e.g.,][]{dzib10}, 
instead of 260 pc, 
then the mass, momentum, and energy increase
further by a factor of 2.5.
Therefore, a main uncertainty on the estimation of the 
outflow parameters comes from the adopted distance to the cloud,
rather than the adopted velocity intervals.

\subsubsection{Mass of Dense Clump}

The HCO$^+$ ($J=4-3$) line emission has a critical density 
of $10^7$ cm$^{-3}$, and thus traces the dense molecular gas.
Here, we estimate the mass and radius of the central
dense clump, using the HCO$^+$ ($J=4-3$) data 
(see Fig. \ref{fig:profile3}a).  
The mass of the HCO$^+$ ($J=4-3$) is evaluated from 
\begin{eqnarray}
M &=& \mu_g m_{\rm H_2} X_{\rm HCO^+}^{-1} \Omega D^2
\sum _i N_{\rm HCO^+,\it i,j} \nonumber \\
&=& 3.1\times 10^{-8} \left({X_{\rm HCO^+} \over 1.2 \times 10^{-9}}\right)^{-1}
\left({D \over 260 {\rm pc}}\right)^2
\left({\Delta \theta \over {\rm arcsec}}\right)^2
 \nonumber \\
&\times& \left({\eta \over 0.57}\right)^{-1}
\left({\sum _{i,j} T_{A,i,j}^* \Delta v \over 
{\rm K \ km \ s^{-1}}}\right)
T_{\rm ex} \exp\left[{43 \ {\rm K} \over T_{\rm ex}}\right] M_\odot
 \ .
\end{eqnarray}
From the above expression, the mass and radius of the central 
clump are estimated to be about 80 $M_\odot$ and 0.1 pc, respectively,
where we assumed LTE and the excitation temperature of 
14 K \citep{andre10}.  
The mean density is estimated to be 
$M_{\rm clump}/(4\pi R_{\rm clump}^3/3) \sim (3-7) \times 10^5$ cm$^{-3}$.
The estimated mean density is smaller than the critical density
of the HCO$^+$ ($J=4-3$) line by an order of magnitude.
This may suggest that the central clump contains multiple small 
subfragments that are not spatially-resolved in our observation.
In fact, recent numerical simulations of cluster formation suggest
that the central dense region of a cluster forming clump 
consists of multiple thin filaments that
tend to fragment into denser blobs by the destruction due to 
the outflow feedback \citep{li10}. However, it is difficult
to verify such a clumpy structure only from our observations.
Higher spatial resolution observations will be needed to reveal 
the density structure of the dense clump.
The fractional abundance of HCO$^+$
relative to H$_2$ is assumed to be $1.2 \times 10^{-9}$, which 
is adopted from \citet{maruta10}.
We note that the fractional abundance of the HCO$^+$ remains uncertain
toward Serpens South. However, from the 1.1 mm data,
the 1.1 mm mass in the same area as that of the HCO$^+$ clump
is evaluated to be $\sim 10^2$ $M_\odot$, which is not far from 
the mass estimated from the HCO$^+$ emission.
Therefore, the adopted HCO$^+$ abundance is likely to be reasonable.
The FWHM one-dimensional velocity width of HCO$^+$ 
is estimated to be $\Delta V \simeq 2.5 - 3.0$ km s$^{-1}$
near the cluster center.  The estimated velocity width corresponds 
to the Mach number of about 5.

\section{Discussion}
\label{sec:discussion}

According to recent theoretical studies, two main 
scenarios for cluster formation are proposed.
One is the rapid, dynamical formation model, in which 
the large-scale flow mainly controls the formation of cluster-forming 
clumps. Star formation is considered to 
complete rapidly within a few crossing times
before initial turbulence decays significantly
\citep{hartmann07}.  
In this model, the magnetic field is expected to be dynamically weak to
promote rapid global gravitational collapse.
To terminate cluster formation, this model
envisions the stellar feedback by the initial star burst that 
disperses the dense gas away from the parent clumps, 
and eventually the clumps are destroyed.

The second scenario is the quasi-equilibrium model.
In this model, star formation process is considered to be
slow and continue at least for several dynamical times
\citep{tan06,li06,matzner07}.
Because supersonic turbulence dissipates quickly in a 
turbulence-crossing time \citep{stone98,maclow99}, 
additional turbulent motions should be injected to maintain 
the supersonic turbulence.
In cluster-forming clumps, protostellar outflow feedback is considered
to be a plausible way of driving a significant fraction of supersonic
turbulence \citep{li06,matzner07,nakamura07,carroll10}. 
The protostellar outflow-driven turbulence can maintain the clump close to 
a quasi-virial equilibrium at least for several dynamical times
\citep{matzner07,nakamura07,carroll10}.
The magnetic field is expected to be dynamically important 
to impede rapid global gravitational collapse.
Very recently, \citet{sugitani11} found that the global magnetic fields
associated with the Serpens South filament are well ordered, 
implying that the moderately-strong magnetic field played an important
role in the formation of the filament. The magnetic structure
is likely to be consistent with the quasi-equilibrium model.

To further constrain these cluster formation models, it is thus 
important to clarify the dynamical state of the clump and the energy 
budget of the current outflows in this region, because 
these two cluster formation models predict the very different star 
formation duration timescales.
In the following, we attempt to estimate the injection rate 
of the turbulent energy due to the outflows and the dynamical 
state of the dense clump.

\subsection{Turbulence Dissipation and Generation}

Following \citet{nakamura11}, we estimate the 
total energy injection rate (mechanical luminosity) due to the 
protostellar outflows and energy dissipation rate of supersonic
turbulence in the dense clump.
From Fig. \ref{fig:bluered}, the mean length of the outflow lobes 
can be estimated to be 0.1 - 0.2 pc, and the mean velocity 
along the line-of-sight direction is about 7 km s$^{-1}$
(see Table \ref{tab:outflow}).  From these values,
the mean dynamical time of the outflows, $t_{\rm dyn}$, can be 
evaluated as $(1-2) \times 10^4$ years, assuming 
$\xi = 57.3^\circ$.  Thus, the total energy injection rate
due to the outflows is obtained as 
\begin{equation}
L_{\rm out} \sim E_{\rm out} / t_{\rm dyn} \sim (0.5 - 1) L_\odot \ .
\end{equation}
On the other hand, 
the energy dissipation rate of supersonic turbulence is estimated to
\begin{equation}
L_{\rm turb} \simeq C \frac{1/2 M\Delta V^2}{\lambda_d/\Delta V}
\end{equation}
where $C (\simeq 0.33)$ is the nondimensional coefficient, $\Delta V$
is the one-dimensional (1D) FWHM velocity width corrected 
for the contribution of the thermal pressure assuming the 
temperature of 14 K (see Equation [3] of \citet{maruta10}), 
and $\lambda_d$ is the driving scale of supersonic turbulence.
From the HCO$^+$ ($J=4-3$) data, we can derive 
the energy dissipation rate as 
\begin{equation}
L_{\rm turb} \simeq 0.1 - 0.3 L_\odot  \ ,
\end{equation} 
where we adopted the velocity width and the driving scale $\lambda_d$
as $\Delta V \sim 2.5 - 3.0$ km s$^{-1}$ and 
the clump diameter of $0.2 - 0.4$ pc, respectively.
Both the energy injection rate $L_{\rm out}$ and 
energy dissipation rate $L_{\rm turb}$ are proportional to the distance
to the cloud, $D$. Therefore, the relative importance 
between $L_{\rm out}$ and $L_{\rm turb}$ is irrespective of the 
distance uncertainty.
The energy injection rate  due to the outflows is somewhat larger 
than the energy dissipation rate.
Therefore, we conclude that the outflow feedback can contribute
significantly to the generation of supersonic turbulence in the clump.

Since we identified about 5 pairs of the outflows in the densest part, 
the mean outflow momentum for a single outflow may be 
estimated to be about 2 M$_\odot$ km s$^{-1}$.
If we assume the median stellar mass of 0.5 M$_\odot$, then
this gives the outflow momentum per unit stellar mass
of about 4 km s$^{-1}$, corresponding
to the non-dimensional parameter $f$ of about 0.04
(see also Maury et al. 2009 for the case of NGC2264-C), which
gauges the strength of a outflow, where the wind velocity of 100 km
s$^{-1}$ is adopted. This value is somewhat smaller than
the fiducial values adopted by \citet{matzner00}, \citet{li06}, 
and \citet{nakamura07}. 
The estimated value may be somewhat underestimated because  
the outflow lobes are assumed to be optically-thin.
If the effect of the optical depth is taken into account,
the value of $f$ is presumably larger than the estimated 
value by a factor of a few.
Even for a small value of $f\sim 0.1$, 
\citet{nakamura07} demonstrated that the outflow feedback can 
supply sufficient momentum to dynamically support a parsec-scale 
cluster-forming clump for several free-fall times, 
by means of the 3D MHD turbulent simulations.
We note that if the distance to Serpens South is as large 
as about 900 pc, the upper limit of the distance to W40
\citep{rodney08}, the outflow strength $f$ is estimated to be 
$f\sim 0.4$.

\subsection{Dynamical State of Dense Clump}

In the following, we investigate the dynamical state of the 
Serpens South clump, by applying the virial analysis.
The virial equation for a sphere is given by 
\begin{equation}
\frac{1}{2} \frac{\partial ^2 I}{\partial t^2} = 2U+W \ ,
\end{equation}
where the terms, $I$, $U$, and $W$, denote the moment of inertia, 
internal kinetic energy, and gravitational energy, respectively.
For simplicity, we neglect the surface pressure term.
A clump is in virial equilibrium when $2U+W=0$.
The terms $U$ and $W$ are expressed as 
\begin{equation}
U=\frac{3M\Delta V^2}{16\ln 2}
\end{equation} 
and
\begin{equation}
W=-a\frac{GM^2}{R} 
\left[1-\left(\frac{\Phi}{\Phi_{\rm cr}}\right)^2\right]  \ , 
\end{equation} 
respectively, where the values $\Phi$ and $\Phi_{\rm cr}$ are, 
respectively, the magnetic flux penetrating the clump 
and the critical magnetic flux above 
which the magnetic field can support the clump against the self-gravity.
Here, we adopt $\Phi=0$ for simplicity, which means no magnetic support.
The value $a$ is a dimensionless parameter of order unity 
which measures the effects of a nonuniform or nonspherical mass 
distribution \citep{bertoldi92}. For a uniform sphere and a 
centrally-condensed sphere with $\rho \propto r^{-2}$, 
$a = 3/5$ and 1, respectively. For the HCO$^+$ clump, 
the effects of the nonspherical mass distribution appear to be 
small because the aspect ratios are not so far from unity 
(see also Figure 2 of Bertoldi \& McKee 1992).
Here, we adopt $a=1$ because the clump seems centrally-condensed.

Then, the terms $U$ and $W$ are estimated to be
$2U=270-390$ M$_\odot$ km$^2$ s$^{-2}$ and $W=-350$ M$_\odot$ km$^2$
s$^{-2}$, respectively, which yields  the virial ratio, 
the ratio between $2U$ and $W$, of 0.8 $-$ 1.1.
This value increases by a factor of about 1.3, 
if we adopt the magnetic field strength derived by \citet{sugitani11}.
The dense clump as a whole is close to quasi-virial equilibrium.
On the other hand, the outflow kinetic energy of the central dense part
detected in the CO ($J=3-2$) emission is derived as  
$E_{\rm out}\sim 24$ M$_\odot$ km$^2$ s$^{-2}$, corresponding to about 7 \% 
the global gravitational energy, where we adopted the value 
corrected for the inclination angle $\xi = 57.3^\circ$. 
Therefore, the energy input due to the current 
outflows may not contribute to change
the global kinetic energy significantly. In other words,
the current outflow activity observed in Serpens South 
appears not to be enough to disperse the whole
cluster-forming clump (see also Maury et al. 2009 for the case of NGC2264-C).
We note that the terms $U$ and $W$ increase with $D^2$ and $D^3$,
respectively, and therefore the virial ratio decreases
by a factor of 1.6 if we adopt $D=410$ pc, instead of 260 pc.
Since this factor and the factor due to the magnetic support
almost cancel each other out, the dense clump is likely to be close to the 
virial equilibrium.

The presence of Class II YSOs in this region suggests that
star formation has lasted over $10^6$ years, assuming the 
typical Class II timescale of about $10^6$ yr.
However, the low fraction of Class II relative to Class I
sources implies that the ages of the YSOs in this region 
may be younger than those in other nearby cluster-forming regions 
\citep{gutermuth09}. 
The age of this region may be $10^{5-6}$ yr.
This corresponds to a few $\sim$ 20 free-fall times of the clump 
assuming the mean density of $5\times 10^5$ cm $^{-3}$, 
where 
\begin{equation}
t_{\rm ff} = \left(\frac{3\pi}{32G\rho_0}\right)^{1/2} \ .
\end{equation}
According to \citet{gutermuth08}, about 30 Class I and II stars are 
associated with the area where we detected the strong HCO$^+$ emission.
Assuming the median stellar mass of 0.5 M$_\odot$, 
the star formation efficiency in the dense part is derived as
about 15 $-$ 20 \%, as reasonably high as those of 
other nearby cluster-forming clumps.
This corresponds to about 1 $-$ 5 \% of the 
star formation rate per free-fall time.  This value 
is in good agreement with that estimated by 
\citet{krumholz07}.
Since the current outflow activity does not provide enough
kinetic energy to destroy the clump, subsequent 
star formation is expected to continue.
This is consistent with the quasi-virial equilibrium 
model for which cluster formation continues for 
a few or more free-fall times.

\section{Summary}
\label{sec:summary}

We carried out the CO ($J=3-2$) and HCO$^+$ ($J=4-3$) observations
toward the nearest cluster-forming IRDC, Serpens South,
using the ASTE 10 m telescope.
The main results are summarized as follows.

1. We found that many outflow components concentrate in the
dense part where the protocluster resides.
Most of these outflow components appear to move away from the 
dense part.
In the northern part, we identified only three bipolar outflows, 
indicating that the star formation is less active in the northern 
part.

2. Most of the outflow components tend to be anti-correlated with
the dense gas traced by the 1.1 mm continuum emission.
Furthermore, the propagation directions of the outflows 
are across the global magnetic field direction
\citep{sugitani11}.
A couple of strong redshifted outflow components appear to 
interact with the dense gas.

3. The central dense clump  detected by the HCO$^+$ ($J=4-3$) emission
has a mass of 80 $M_\odot$ and a radius of 
about 0.1 pc, with an axis ratio of about 2.  The average density and velocity
width are estimated at about $5\times 10^5$ cm$^{-3}$ and 
$2.5-3$ km s$^{-1}$, respectively.  
The dense clump has very clumpy structures. Some of the subfragments 
appear to coincide with the positions of the Class 0 candidates 
identified by \citet{bontemps10}.

4. We estimated the global physical quantities of the outflows.
The total outflow mass, momentum, and energy seem smaller than those 
of the Serpens Cloud Core, a nearby typical parsec-scale cluster-forming
clump, located about 3$^\circ$ north of Serpens South.  
However, the characteristic outflow speed appears somewhat
larger than that of the Serpens Cloud Core. This may imply that 
the YSO populations of Serpens South are younger than those of 
the Serpens Cloud Core.

5. The outflow energy injection rate is likely to be 
somewhat larger than the energy dissipation rate 
of the supersonic turbulence, suggesting 
that the outflow feedback can significantly contribute to 
the generation of the supersonic turbulence in the dense clump.
Assuming the median stellar mass of 0.5 M$_\odot$,
the mean outflow momentum per unit stellar mass
is estimated to be about 4 km s$^{-1}$, under
the assumption of optically-thin gas.
This mean outflow momentum corresponds to the non-dimensional 
outflow parameter of $f \sim 0.04$, which gauges 
the strength of a outflow. 
This value of $f$ is somewhat smaller than the fiducial values 
of $f=0.4$ adopted by \citet{matzner00}, \citet{li06}, and \citet{nakamura07}.
If we take into account the effect of the optical depth,
the outflow strength $f$ presumably increases by a factor of a few.

6. The total outflow energy appears significantly smaller than 
the global gravitational energy of the dense part where
the protocluster is located.  In other words, it may be 
difficult to destroy the cluster-forming clump by
the current outflow activity.  This may be inconsistent
with the dynamical model of cluster formation, for which
the outflow feedback due to the initial star burst is envisioned
to disperse the dense gas from the cluster-forming clump.

\acknowledgments 
This work is supported in part by a Grant-in-Aid for Scientific Research
of Japan (20403003, 20540228, 22340040) and 
National Science Council of Taiwan (Grant No. 98-2112-M-001-002-MY3). 
We thank Philippe Andr{\'e} for helpful comments.
We are grateful to the ASTE staffs for both operating the ASTE telescope and 
helping us with the data reduction.
The ASTE project is driven by Nobeyama Radio Observatory 
(NRO), a branch of National Astronomical Observatory of Japan (NAOJ), 
in collaboration with University of Chile, and Japanese institutes 
including University of Tokyo, Nagoya University, 
Osaka Prefecture University, Ibaraki University, and Hokkaido
University.
The observations with ASTE were carried out remotely from Japan by using
NTT's GEMnet2 and its partner R\&E (Research and Education) networks,
which are based on AccessNova collaboration of University of Chile, NTT
Laboratories, and National Astronomical Observatory of Japan.
This research has made use of the NASA/IPAC Infrared Science Archive,
which is operated by the Jet Propulsion Laboratory, California Institute
of Technology, under contract with the National Aeronautics and 
Space Administration.

\clearpage

\begin{deluxetable}{llll}
\tabletypesize{\scriptsize}
\tablecolumns{4}
\tablecaption{Characteristics of Identified CO ($J=3-2$) Outflow Lobes}
\tablewidth{\columnwidth}
\tablehead{\colhead{Name} 
& \colhead{counterpart\tablenotemark{a}} 
& \colhead{driving source\tablenotemark{b}} 
& \colhead{features} 
}
\startdata
B1 & unknown & source 1  & central part \\
B2 & unknown & unknown & central part \\
B3 & unknown & unknown & central part \\
B4 & R2, R3 & source 2 & central part, strong \\
B5 & unknown & source 4 & central part, related to R4 \\
B6 & unknown & source 3 & central part \\
B7 & no & unknown & southern part \\
B8 & R2, R3 ? & source 2 ? & central part, a head of B4 ?\\
B9 & unknown & source 3 ? & a head of B6 ?, outflow or infall gas \\
B10 & R5 & unknown & northern part, clear bipolar \\
B11 & R6 & unknown & northern part, clear bipolar \\
B12 & R7 & unknown & northern part, clear bipolar \\
B13 & no & unknown & eastern part \\
B14 & no & source 5 & southern part \\
B15 & R8 & source 6 & southern part, clear bipolar \\
B16 & R10 & Herschel source & northern part, clear bipolar \\
R1 & unknown & source 1 & central part \\
R2 & B4 & source 2 & central part, strong  \\
R3 & B4 & source 2 & central part, strong \\
R4 & unknown & source 4 & central part, related to B5 \\
R5 & B10 & unknown  & northern part, clear bipolar \\
R6 & B11 & unknown & northern part, clear bipolar \\
R7 & B12 & unknown & northern part, clear bipolar \\
R8 & B15 & source 6 & southern part, clear bipolar \\
R9 & no & unknown & northern part, \\
R10 & B16 & Herschel source & northern part, clear bipolar
\enddata
\tablenotetext{a}{Possible counterparts are indicated. ``no'' means
that only one component is seen.}
\tablenotetext{b}{Possible driving sources are indicated. 
Sources 1  through 6 are indicated in 
Figs. \ref{fig:spitzer2} and \ref{fig:spitzer3}.}
\tablecomments{See Section \ref{subsec:outflow} in detail.}
\label{tab:outflow lobe}
\end{deluxetable}

\begin{deluxetable}{lllll}
\tabletypesize{\scriptsize}
\tablecolumns{5}
\tablecaption{Global outflow properties derived from CO ($J=3-2$)}
\tablewidth{\columnwidth}
\tablehead{\colhead{} 
& \colhead{Mass ($M_\odot$)} 
&\colhead{Momentum ($M_\odot$ km s$^{-1}$)} 
&\colhead{Energy ($M_\odot$ km$^2$ s$^{-2}$)} 
&\colhead{$V_{\rm out}$ (km s$^{-1}$)} 
}
\startdata
Blue & 0.34 & 3.7 & 25.9 & 10.9 \\
Red & 0.27 & 3.9 & 38.7 & 14.4 \\
Total & 0.61  & 7.6  & 64.6 & 12.5
\enddata
\label{tab:outflow}
\end{deluxetable}

\clearpage


\begin{thebibliography}{99}
\bibitem[Allen et al.(2006)]{allen06}
Allen, L. et al. 2006. in Protostars and Planets V, eds. B. Reipurth,
		D. Jewitt, and K. Keil (The University of Arizona
		Press), p. 361
\bibitem[Andr\'e et al.(2010)]{andre10}
Andr\'e, P. et al. 2010, \aap, 518, L102
\bibitem[Andr\'e et al.(2007)]{andre07}
Andr\'e, P. et al. 2007, \aap, 472, 519
\bibitem[Arce et al.(2010)]{arce10}
Arce, H. G., Borkin, M. A., Goodman, A. A., Pineda, J. E., \& Halle,
		M. W. 2010, \apj in press
\bibitem[Bertoldi \& McKee(1992)]{bertoldi92}
Bertoldi, F., \& McKee, C. F. 1992, \apj, 395, 140
\bibitem[Bontemps et al.(1996)]{bontemps96}
Bontemps, S., Andr\'e, P., Terebey, S., \& Cabrit, S. 1996, \aap, 311, 858
\bibitem[Bontemps et al.(2010)]{bontemps10}
Bontemps, S. et al. 2010, \aap, 518, L85
\bibitem[Boogert et al.(2002)]{boogert02}
Boogert, A. C. A., Hogerheijde, M. R., Ceccarelli, C., Tielens,
		A. G. G. M., van Dishoeck, E. F., Blake, G. A., Latter,
		W. B., \& Motte, F., 2002, \apj, 570, 708
\bibitem[Butler \& Tan(2009)]{butler09}
Butler, M. J., Tan, J. C. 2009, \apj, 696, 484
\bibitem[Carroll et al.(2010)]{carroll10}
Carroll, J. J., Frank, A., \& Blackman, E. G. 2010, \apj, 722, 145

\bibitem[Cabrit \& Bertout(1992)]{cabrit92}
Cabrit, S. \& Bertout, C. 1992, \aap, 261, 274
\bibitem[Davis et al.(1999)]{davis99}
Davis, C. J., Matthews, H. E., Ray, T. P., Dent, W. R. F., \& Richer,
		J. S. \mnras, 309, 141
\bibitem[Dzib et al.(2010)]{dzib10}
Dzib, S. et al. 2010, \apj, 718, 610
\bibitem[Ezawa et al.(2004)]{ezawa04}
Ezawa, H., Kawabe, R., Kohno, K., \& Yamamoto, S. 2004, Proc. SPIE, 5489, 763
\bibitem[Elmegreen(2007)]{elmegreen07}
Elmegreen, B. G. 2007, \apj, 668, 1064
\bibitem[Graves et al.(2010)]{graves10}
Graves,, S. F., et al. 2010, \apj, in press (arXiv:1006.0891)
\bibitem[Gorlova et al.(2010)]{gorlova10}
Gorlova, N., Steinhauer, A., \& Lada, E. 2010, \apj, 716, 634
\bibitem[Gutermuth et al.(2008)]{gutermuth08}
Gutermuth, R. A. et al. \apj, 2008, 673, L151
\bibitem[Gutermuth et al.(2009)]{gutermuth09}
Gutermuth, R. A. et al. \apjs, 2009,
\bibitem[Gutermuth et al.(2011)]{gutermuth11}
Gutermuth, R. A. et al. 2011, submitted to \apj
\bibitem[Hartmann \& Burkert(2007)]{hartmann07}
Hartmann, L., \& Burkert, A. 2007, \apj, 654, 988
\bibitem[Kamazaki et al.(2003)]{kamazaki03}
Kamazaki, T., Saito, M., Hirano, N., Umemoto, T. \& Kawabe, R. \apj,
		2003, 584, 357
\bibitem[Kirk et al.(2007)]{kirk07}
Kirk, H. et al. 2007, \apj, 668, 1042
\bibitem[Krumholz et al.(2006)]{krumholz06}
Krumholz, M. R., Matzner, C. D., McKee, C. F. 2006, \apj, 653, 361
\bibitem[Krumholz \& Tan(2007)]{krumholz07}
Krumholz, M. R. \& Tan, J. C. 2007, \apj, 656, 959
\bibitem[Klessen \& Hennebelle(2010)]{klessen10}
Klessen, R. S. \& Hennebelle, P. 2010, \aap, 520, 17
\bibitem[Lada \& Lada(2003)]{lada03}
Lada, C. J. \& Lada, E. A. \araa, 2003, 41, 57
\bibitem[Li \& Nakamura(2006)]{li06}
Li, Z.-Y. \& Nakamura, F. \apj,	2006, 640, L187
\bibitem[Li et al.(2010)]{li10}
Li, Z.-Y., Wang, P., Abel, T., \& Nakamura, F. \apj, 2010, 720, L26
\bibitem[Mac Low(1999)]{maclow99}
Mac Low, M.-M. 1999, \apj, 524, 169
\bibitem[Maruta et al.(2010)]{maruta10}
 Maruta, H., Nakamura, F., Nishi, R., Ikeda, N., \& Kitamura, Y. 2010,
		\apj, 714, 680
\bibitem[Matzner \& McKee(2000)]{matzner00}
Matzner, C. D., \& McKee, C. F. 2000, \apj, 545, 364
\bibitem[Matzner(2007)]{matzner07}
Matzner, C. D. \apj, 2007, 659, 1394
\bibitem[Maury et al.(2009)]{maury09}
Maury A., Andr\'e, P., \& Li, Z.-Y. \apj, 2009, 499, 175
\bibitem[McKee \& Ostriker(2007)]{mckee07}
McKee, C. F., \& Ostriker, E. C. 2007, \araa, 45, 565
\bibitem[Myers(2009)]{myers09}
Myers, P. C. 2009, \apj, 700, 1609
\bibitem[Nakamura \& Li(2007)]{nakamura07}
Nakamura, F. \& Li, Z.-Y. 2007, \apj,	662, 395
\bibitem[Nakamura \& Li(2011)]{nakamura11b}
Nakamura, F. \& Li, Z.-Y. 2011, submitted to \apj
\bibitem[Nakamura et al.(2011)]{nakamura11}
Nakamura, F. et al. 2011, \apj, 726, 46
\bibitem[Noriega-Crespo et al.(2004)]{noriega-crespo04}
Noriega-Crespo, A., Moro-Martin, A., Carey, S., 
Morris, P. W., Padgett, D. L., Latter, William B., \& Muzerolle, J.
2004, \apjs, 154, 402
\bibitem[Norman \& Silk(1980)]{norman80}
Norman, C., \& Silk, J. 1980, \apj, 238, 158
\bibitem[Peters et al.(2010)]{peters10} 
 Peters, T., Klessen, R. S., Mac Low, M.-M., Banerjee, R.,
2010, \apj, 725, 134 
\bibitem[Perreto et al.(2006)]{perreto06}
Perreto, N. et al. 2006, \aap, 445, 979
\bibitem[Perreto \& Fuller(2009)]{perreto09}
Perreto, N. \& Fuller, G. A. 2009, \aap, 505, 405 
\bibitem[Rathborne et al.(2006)]{rathborne06}
Rathborne, J. M., Jackson, J. M., Simon, R. 2006, \aap, 641, 389
\bibitem[Ridge et al.(2003)]{ridge03}
 Ridge, N. A., Wilson, T. L., Megeath, S. T., Allen, L. E., \&  Myers,
		P. C. 2003, \aj, 126, 286
\bibitem[Rodney \& Reipurth(2008)]{rodney08}
Rodney, S. A. \& Reipurth, B. 2008, in Handbook of Star Forming Regions
		Vol. II, Astronomical Society of the Pacific, eds. Bo Reipurth
\bibitem[Saito et al. (2008)]{saito08}
Saito, H., Saito, M., Yonekura, Y., \& Nakamura, F. 2008, \apjs, 178, 302
\bibitem[Sandell \& Knee(2001)]{sandell01}
Sandell, G. \& Knee, L. B. G. 2001, \apj, 546, L49
\bibitem[Schneider et al.(2010)]{schneider10}
Schneider, N., Csengeri, T., Bontemps, S., Motte, F., Simon, R.,
		Hennebelle, P., Federrath, C., \& Klessen, R. \aap,
		2010, 520, 49
\bibitem[Stone et al.(1998)]{stone98}
Stone, J. M., Ostriker, E. C., \& Gammie, C. F. 1998, \apj, 508, L99
\bibitem[Straizys et al.(1996)]{straizys96}
Straizys, V., Cernis, K., 
\& Bartasiute, S. 1996, Baltic Astron., 5, 125
\bibitem[Sugitani et al.(2010)]{sugitani10}
Sugitani, K. et al. 2010,  \apj, 716, 299
\bibitem[Sugitani et al.(2011)]{sugitani11}
Sugitani, K. et al. 2011,  \apj, in press (arXiv:1104.2977)
\bibitem[Takami et al.(2010)]{takami10}
Takami, M., Karr, J. L., Koh, H., Chen, H.-H., \& Lee, H.-T. 
\apj, 2010,  \apj, 720, 155
\bibitem[Tan et al.(2006)]{tan06}
Tan, J. C., Krumholz, M. R., \& McKee, C. F. 2006, \apj, 641, L121
\bibitem[Vazquez-Semadeni et al.(2010)]{enrique10}
Vazquez-Semadeni, E., Colin, P., Gomez, G. C., 
Ballesteros-Paredes, J., Watson, A W. 2010, \apj, 715, 1302
\bibitem[Walsh et al.(2007)]{walsh07}
Walsh, A. J., Myers, P. C., Di Francesco, J., Mohanty, S., 
Bourke, T. L., Gutermuth, R., Wilner, D. 2007, \apj, 655, 958
\bibitem[Wang et al.(2010)]{wang10}
Wang, P., Li, Z.-Y., Abel, T., \& Nakamura, F. 2010, \apj, 709, 27
\bibitem[Wilson et al.(2008)]{wilson08}
Wilson, G. W., et al. 2008, \mnras, 386, 807
\bibitem[Winston et al.(2010)]{winston10}
Winston, E. et al. 2010, \aj, 140, 266
\bibitem[Zhang \& Wang(2009)]{zhang09}
Zhang, M. \& Wang, H. 2009, \aj, 138, 1830
\end{thebibliography}
\end{document}